\documentclass[conference]{IEEEtran}
\IEEEoverridecommandlockouts
\usepackage{acronym}
\acrodef{leo}[LEO]{low Earth orbit}
\acrodef{meo}[MEO]{medium Earth orbit}
\acrodef{geo}[GEO]{geostationary Earth orbit}
\acrodef{ut}[UT]{user terminal}
\acrodef{upa}[UPA]{uniform planar array}
\acrodef{csi}[CSI]{channel state information}
\acrodef{ofdm}[OFDM]{orthogonal frequency division multiplexing}
\acrodef{los}[LoS]{line-of-sight}
\acrodef{toa}[TOA]{time-of-arrival}
\acrodef{pace}[PACE]{positioning-aided channel estimation}
\acrodef{mcrb}[MCRB]{misspecified Cramér-Rao bound}
\acrodef{awgn}[AWGN]{additive white Gaussian noise}
\acrodef{crb}[CRB]{Cramér-Rao bound}
\acrodef{cfo}[CFO]{carrier frequency offset}
\acrodef{peb}[PEB]{position error bound}
\acrodef{crb}[CRB]{Cramér-Rao bound}
\acrodef{lb}[LB]{lower bound}
\acrodef{rmse}[RMSE]{root mean squared error}
\acrodef{fim}[FIM]{Fisher information matrix}
\acrodef{tdd}[TDD]{time division duplex}
\acrodef{wmmse}[WMMSE]{weighted minimal mean squared error}
\acrodef{fp}[FP]{fractional programming}
\acrodef{qcqp}[QCQP]{quadratically constrained quadratic program}
\acrodef{mrt}[MRT]{maximum ratio transmission}
\acrodef{gnss}[GNSS]{global navigation satellite system}
\acrodef{itu}[ITU]{International Telecommunication Union}
\acrodef{pab}[PAB]{Position-Aided Beamforming}
\acrodef{vdb}[VDB]{Vertically Directed Beamforming}
\acrodef{bs}[BS]{base station}
\acrodef{ntn}[NTN]{non-terrestrial network}
\acrodef{uav}[UAV]{unmanned aerial vehicle}
\acrodef{tdma}[TDMA]{time-division multiple access}
\acrodef{haps}[HAPS]{high-altitude platform stations}
\acrodef{3gpp}[3GPP]{3rd Generation Partnership Project}
\acrodef{nr}[NR]{New Radio}
\acrodef{dof}[DoF]{degree of freedom}
\acrodefplural{dof}[DoFs]{degrees of freedom}
\acrodef{6g}[6G]{sixth generation}
\acrodef{bse}[BSE]{beam squint effect}
\acrodef{cp}[CP]{cyclic prefix}
\acrodef{elaa}[ELAA]{extremely large antenna array}
\acrodef{ff}[FF]{far-field}
\acrodef{las}[L\&S]{localization and sensing}
\acrodef{nf}[NF]{near-field}
\acrodef{ris}[RIS]{reconfigurable intelligent surface}
\acrodef{rtt}[RTT]{round-trip-time}
\acrodef{sinr}[SINR]{signal-to-interference-plus-noise ratio}
\acrodef{sns}[SNS]{spatial non-stationarity}
\acrodef{swm}[SWM]{spherical wave model}
\acrodef{soc}[SOC]{second-order cone}
\acrodef{socp}[SOCP]{second-order cone program}
\acrodef{siso}[SISO]{single-input-single-output}
\acrodef{mimo}[MIMO]{multiple-input multiple-output}
\acrodef{ue}[UE]{user equipment}
\acrodef{dmimo}[D-MIMO]{distributed MIMO}
\acrodef{sp}[SP]{scatter point}
\acrodef{nlos}[NLoS]{non-line-of-sight}
\acrodef{tdoa}[TDOA]{time-difference-of-arrival}
\acrodef{am}[AM]{artificial multipath}
\acrodef{an}[AN]{artificial noise}
\acrodef{psd}[PSD]{power spectral density}
\acrodef{pdf}[PDF]{probability distribution function}
\acrodef{aoa}[AOA]{angle-of-arrival}
\acrodef{aod}[AoD]{angle of departure}
\acrodef{moo}[MOO]{multi-objective optimization}
\acrodef{qos}[QoS]{quality of service}
\acrodef{sdp}[SDP]{semi-definite programming}
\acrodef{lmi}[LMI]{linear matrix inequality}
\acrodef{sdr}[SDR]{semi-definite relaxation}
\acrodef{rcs}[RCS]{radar cross section}
\acrodef{isac}[ISAC]{integrated sensing and communication}
\acrodef{pdd}[PDD]{penalty dual decomposition}
\acrodef{bcd}[BCD]{block coordinate descent}
\acrodef{minlp}[MINLP]{mixed-integer nonlinear programming}
\acrodef{isl}[ISL]{inter-satellite link}
\acrodef{ais}[AIS]{automatic identification system}
\acrodef{rfc}[RFC]{radio frequency chain}
\acrodef{bbm}[BBM]{break-before-make}
\acrodef{mbb}[MBB]{make-before-break}
\acrodef{iui}[IUI]{inter-user interference}

\usepackage{pgfplots}
\pgfplotsset{compat=1.18}
\usepackage{tikz}
\usetikzlibrary{calc,patterns}
\makeatletter
\newcommand{\gettikzxy}[3]{%
  \tikz@scan@one@point\pgfutil@firstofone#1\relax
  \edef#2{\the\pgf@x}%
  \edef#3{\the\pgf@y}%
}
\usetikzlibrary{spy,backgrounds,arrows.meta,decorations.pathreplacing,decorations.pathmorphing}
\usepackage{color}
\usepackage{mathrsfs}
\usepackage{booktabs}
\usepackage{amsmath}
\usepackage{graphicx}
\usepackage{epstopdf}
\usepackage{amssymb}
\usepackage{amsfonts}
\usepackage{amsthm}
\usepackage{cite}
\usepackage{bm,comment}
\usepackage{algorithm}
\usepackage{algpseudocode}

\usepackage{subeqnarray}
\usepackage{subfigure}
\usepackage{multicol}
\usepackage{multirow}
\usepackage{diagbox}
\usepackage{slashbox}
\usepackage{stfloats}
\usepackage{float}
\usepackage{color}
\usepackage{cases}
\usepackage{lipsum}
\newtheoremstyle{zackplain}{\topsep}{\topsep}{\itshape}{}{\bfseries}{:}{.5em}{}
\newtheoremstyle{zackremark}{0pt}{0pt}{\normalfont}{\parindent}{\bfseries}{:}{.5em}{\thmname{#1}\thmnumber{ #2}\thmnote{ (#3)}}
\theoremstyle{zackplain}

\theoremstyle{zackremark}
\newtheorem{remark}{Remark}
\usepackage[hidelinks]{hyperref}
\usepackage{geometry}
\allowdisplaybreaks
\begin{document}
\AddToHookNext{shipout/foreground}{\put(0.5\paperwidth,-0.32in){\makebox[0pt]{%
  \footnotesize\begin{tabular}[t]{@{}c@{}}This work has been submitted to the IEEE for possible publication.\\ Copyright may be transferred without notice, after which this version may no longer be accessible.\end{tabular}}}}
\setlength{\textfloatsep}{4pt}
\setlength{\dbltextfloatsep}{8pt plus 2pt minus 2pt}
\setlength{\parskip}{0pt plus 2pt}

\bstctlcite{IEEEexample:BSTcontrol}
\title{Handover-Aware Power Minimization for \\
 Networked LEO Satellite Communications: \\
Joint Cooperative Beamforming and Scheduling}
\author{
Yuchen Zhang, \emph{Member, IEEE}, Eva Lagunas, \emph{Senior Member, IEEE}, \\Symeon Chatzinotas, \emph{Fellow, IEEE}, and Tareq Y. Al-Naffouri, \emph{Fellow, IEEE}
\vspace{-3mm}
\thanks{
This publication is based upon work supported by King Abdullah University of Science and Technology (KAUST) under Award No. ORFS-CRG12-2024-6478 and Global Fellowship Program under Award No. RFS-2025-6844.

Yuchen Zhang and Tareq Y. Al-Naffouri are with the Electrical and Computer Engineering Program, Computer, Electrical and Mathematical Sciences and Engineering (CEMSE), King Abdullah University of Science and Technology (KAUST), Thuwal 23955-6900, Kingdom of Saudi Arabia (e-mail: \{yuchen.zhang; tareq.alnaffouri\}@kaust.edu.sa).

Eva Lagunas and Symeon Chatzinotas are with the Interdisciplinary Centre for Security, Reliability and Trust (SnT), University of Luxembourg, 1855 Luxembourg City, Luxembourg (e-mail: \{eva.lagunas; Symeon.Chatzinotas\}@uni.lu).
}}
\maketitle

\begin{abstract}
Networked low Earth orbit (LEO) satellite constellations enabled by inter-satellite links offer a promising path toward ubiquitous broadband non-terrestrial services, as cooperative transmission increases the received power and suppresses interference. However, fast orbital motion induces frequent handovers, which cost energy and transmission time, while the limited number of radio-frequency chains per satellite couples user scheduling with cooperative beamforming. In this paper, we investigate handover-aware power-efficient downlink transmission in networked LEO systems under statistical channel state information. Specifically, we first introduce a two-segment frame structure that separates the handover procedure from user-plane transmission, and propose a power consumption model that captures the energy to establish each new satellite-user link and the transmission time lost during its handover. We then formulate a per-frame network-wide power minimization problem that jointly optimizes cooperative beamforming and user scheduling, and develop an iterative algorithm that combines a reweighted $\ell_2$ surrogate with a penalty-based relaxation and a fractional-programming inner loop, which yields convex subproblems.
Simulation results with Red Sea ship traffic demonstrate that, at $25$~Mbps per user, the proposed design needs 14\% less power than distance-based cooperative scheduling and 15\% less than a joint design that ignores the established links, while non-cooperative transmission meets the rate requirements in only 14\% of the frames.

\end{abstract}


\IEEEpeerreviewmaketitle

\section{Introduction}
\Acp{ntn} are recognized as a key component of \ac{6g} systems for providing ubiquitous \emph{broadband} connectivity and bridging the digital divide, especially over sparsely populated and hard-to-reach areas~\cite{imt2030vision,6GtakeShape,konstantin2025proc,eva2025standard}. Among \ac{ntn} platforms, \ac{leo} satellite constellations have attracted substantial attention from both academia and industry~\cite{konstantin2025proc}, since their lower orbital altitudes reduce the propagation latency and the path loss compared with \ac{meo} and \ac{geo} systems. In addition, \ac{leo} satellites can carry large antenna arrays, which enable multi-user beamforming for high-throughput services~\cite{you2020jsac,you2024twc}.

Despite these advantages, the long propagation distance and the limited on-board power restrict the rate that a single \ac{leo} satellite can provide to a user. This motivates \emph{multi-satellite cooperation} instead of conventional \emph{single-satellite} service. With \acp{isl}~\cite{halim2021vtm}, multiple \ac{leo} satellites can be interconnected to enable \emph{networked \ac{leo} cooperation}, where several satellites jointly serve users via cooperative beamforming~\cite{zack2025twc,yafei2026jsac,zack2026wcm,yafei2026tcom,zack2026tcom,kexin2024twc,asynLEO2024TWC,zack2025decentral}. This cooperation can be viewed as a non-terrestrial counterpart of terrestrial cell-free massive \ac{mimo}, coordinating distributed transmitters to exploit their additional spatial \acp{dof} for interference mitigation and a higher received signal power~\cite{halim2022oj,halim2023oj,moewin2025jsac,meixia2024twc}.

There has been a growing body of work on networked \ac{leo} cooperative transmission, addressing cooperative beamforming, resource allocation, and designs based on statistical \ac{csi} under \ac{leo}-specific channel characteristics~\cite{zack2025twc,yafei2026jsac,zack2025decentral,moewin2025jsac,asynLEO2024TWC}. However, most of these works maximize a rate metric, e.g., the sum rate or the minimum rate, while a more practically relevant question in \ac{leo} systems is how to \emph{minimize the power consumption subject to prescribed service rates}~\cite{poor2024tsp}. This is because the on-board power supply of \ac{leo} satellites is limited~\cite{bruno2024twc}. Moreover, each satellite has a limited number of \acp{rfc}, so it can serve only a small subset of the \acp{ut} simultaneously, which necessitates \emph{user scheduling}. Yet, in many existing works, scheduling is either ignored or handled in a decoupled manner, e.g., by fixing the schedule before optimizing the beamformers~\cite{zack2025twc,zack2025decentral}. Such a decoupled design is suboptimal in general, because the schedule determines which satellites cooperate for each \ac{ut} and which \acp{ut} interfere with each other. Hybrid beamforming and user scheduling with limited \acp{rfc} were designed jointly in~\cite{meixia2024twc} to increase the spectral efficiency of multi-satellite cooperation, but without accounting for handovers, which we discuss next.

While networked \ac{leo} cooperation offers significant performance gains, the orbital dynamics introduce an additional challenge that is far less pronounced in terrestrial systems. \ac{leo} satellites move at several kilometers per second, so the satellite-\ac{ut} distances change continuously, and the set of visible satellites changes within minutes. As a result, the satellite-\ac{ut} links must be updated frequently, and each new link requires a \emph{handover} procedure, which incurs signaling and energy overhead~\cite{juan2022access}, consumes dedicated radio resources, and may interrupt user-plane transmission~\cite{3gpp.38.300,juan2022access}. This issue becomes more intricate in networked \ac{leo} cooperation: since each \ac{ut} is served by several satellites, a scheduling update may establish several satellite-\ac{ut} links at once, each with its own \emph{handover-related power cost}. Without accounting for this cost, a design underestimates the power consumption. Moreover, it cannot weigh the two options for a \ac{ut} whose satellite recedes: keeping the link, which needs more radiated power as the satellite-\ac{ut} distance grows, or establishing a new link to a closer satellite, which costs a handover.

\looseness=-1 Handover in \acp{ntn} has been addressed at the protocol level, e.g., by the procedures of \ac{3gpp} \ac{nr}-\ac{ntn}~\cite[Sec.~16.14.3]{3gpp.38.300}. Recent studies of \ac{leo} systems have improved the handover performance, e.g., by predictive handovers that cope with long propagation delays~\cite{molisch2024twc}, by geometry- or gain-aware handover triggering and access selection~\cite{juan2022access,tong2025taes}, and by transmission/beamforming adaptations or multi-orbit joint service that mitigate the handover-induced performance loss~\cite{zhiyong2023wcl,han2025twc}. Multi-connectivity can enhance service continuity in \acp{ntn}~\cite{majamaa2024mcom}, and at the \ac{leo} \ac{ut}, low-cost receive beamforming was optimized for make-before-break handover through dual connectivity~\cite{lema2026tcom}. Closer to our setting, power allocation and handover management have been optimized jointly in distributed massive \ac{mimo} \ac{leo} networks~\cite{halim2022oj}, and dynamic satellite clusters have been formed to reduce handovers~\cite{moewin2025jsac}. However, neither work accounts for the power spent on establishing each new satellite-\ac{ut} link or for the transmission time lost during a handover. Consequently, \emph{how to minimize the network-wide power consumption, including this handover cost, through joint cooperative beamforming and user scheduling in networked \ac{leo} systems} remains largely open.

Motivated by the above gaps, this paper develops a handover-aware power minimization framework for networked \ac{leo} satellite downlinks with joint cooperative beamforming and user scheduling. Since the channel changes during the round-trip delay between the satellites and the \acp{ut}, instantaneous \ac{csi} is hard to obtain at the satellites, which motivates performance evaluation and optimization based on statistical-\ac{csi} metrics~\cite{you2020jsac,zack2025decentral,moewin2025jsac}. Accordingly, we adopt the hardening bound, a closed-form lower bound on the ergodic downlink rate, which leads to per-\ac{ut} \ac{qos} constraints that depend only on statistical channel parameters~\cite{Marzetta2016mMIMO,caire2018twc}. The main contributions are summarized as follows.
\begin{itemize}
    \item \textbf{Handover-aware power consumption modeling:}
    We introduce a two-segment frame structure that separates the handover procedure from the user-plane transmission, and propose a power consumption model that captures (i) the handover-related power of each new satellite-\ac{ut} link and (ii) the transmission time that a new link loses during its handover, so that the power consumption of a frame depends on the previous-frame schedule.

    \item \textbf{Handover-aware joint power minimization and algorithmic solution:}
    \looseness=-1 Based on this power model and on the hardening bound, we formulate a per-frame network-wide power minimization problem that jointly optimizes cooperative beamforming and the \ac{leo}-\ac{ut} schedule, which is implicitly defined by the nonzero beamformers, subject to per-\ac{ut} rate constraints that account for the handover time, per-satellite radiated power budgets, and per-satellite limits on the number of served \acp{ut}. To handle the combinatorial scheduling and the nonconvex fractional rate constraints, we develop an iterative algorithm that (i) promotes sparsity in the beamformers of each satellite via a reweighted $\ell_2$ surrogate with a penalty-based relaxation, and (ii) handles the rate constraints by a \ac{fp}-based inner loop, which leads to a sequence of convex subproblems solvable by standard tools.

    \item \textbf{Evaluation with orbital dynamics and maritime user data:}
    \looseness=0 We conduct Monte-Carlo simulations over two-minute trajectories of a Walker-Delta constellation, in which the serving set is re-selected in every frame, for vessels drawn from the ship traffic of the Red Sea. The results
show that the proposed design reduces the network-wide power consumption compared with cooperative schemes with pre-fixed schedules, a non-cooperative scheme with optimized beamforming, and a joint scheduling and beamforming design that ignores the established links, and that it meets the rate requirements in more frames than the pre-fixed schedules when the requirements are high or the \acp{rfc} are few. They also reveal that keeping links as long as possible does not minimize the power, that cooperation makes the proposed design almost insensitive to the handover duration, and that a look-ahead variant, which spreads the handover energy of a new link over the frames in which the link can serve, saves further power.

\end{itemize}

The rest of this paper is organized as follows. Section~\ref{sysmod} presents the signal and channel models. Section~\ref{sec_handover} introduces the frame structure, the handover-aware power consumption model, and the statistical rate metric. Section~\ref{sec_method} formulates the handover-aware joint optimization problem and develops the proposed iterative algorithm. Section~\ref{sec_num} evaluates the proposed design over satellite trajectories and compares it with benchmark schemes. Finally, Section~\ref{sec_conc} concludes the paper.

\emph{Notations:} Vectors and matrices are denoted by bold lowercase and bold uppercase letters, respectively. The Euclidean norm of a vector $\mathbf{a}$ is $\|\mathbf{a}\|$, and $(\cdot)^{\mathsf{T}}$, $(\cdot)^{\mathsf{H}}$, and $(\cdot)^{*}$ denote the transpose, Hermitian transpose, and complex conjugate. The diagonal matrix with diagonal entries $\mathbf{a}$ is $\mathrm{diag}(\mathbf{a})$, the block-diagonal matrix with blocks $\mathbf{A}_1,\ldots,\mathbf{A}_N$ is $\mathrm{blkdiag}(\mathbf{A}_1,\ldots,\mathbf{A}_N)$, and $\mathbf{I}_U$ is the $U\times U$ identity matrix. The real and imaginary parts of a complex scalar $a$ are $\Re\{a\}$ and $\Im\{a\}$, $\jmath=\sqrt{-1}$ is the imaginary unit, and $\otimes$ is the Kronecker product. The expectation and the variance are denoted by $\mathbb{E}[\cdot]$ and $\mathbb{V}[\cdot]$, and the indicator function $1_{+}\{x\}$ equals $1$ if $x>0$ and $0$ otherwise. Finally, $\mathcal{CN}(\mathbf{a},\mathbf{C})$ denotes a circularly symmetric complex Gaussian distribution with mean $\mathbf{a}$ and covariance matrix $\mathbf{C}$.

\begin{figure}[t]
		\centering
		\input{Figures/system_figure.tex}
    \caption{Networked \ac{leo} system in the $k$-th frame. The satellites of the serving set $\mathcal{L}^{(k)}$ cooperate over \acp{isl}, and each serves at most $U_{\max}$ \acp{ut} through its planar array (inset). As one satellite joins the set and another leaves it, UT~3 keeps a persisting link, whereas UT~4 is handed over to a new link.}
		\label{system_fig}
\end{figure}

\section{Signal and Channel Models\label{sysmod}}
As illustrated in Fig.~\ref{system_fig}, we consider cooperative downlink communications in a \emph{networked} \ac{leo} satellite system. A set of $L$ \ac{leo} satellites, called the serving set, jointly serves $U$ \acp{ut}. The satellites are regenerative, i.e., they process the signals on board instead of only relaying them. Each \ac{ut} is modeled as a single-antenna receiver with an omnidirectional (or quasi-omnidirectional) pattern, which lets it receive the signals of several satellites at once, as for the vessels in Fig.~\ref{system_fig}. Each satellite employs a half-wavelength-spaced \ac{upa} comprising
$N = N_{\rm{h}} \times N_{\rm{v}}$ antenna elements, where $N_{\rm{h}}$ and $N_{\rm{v}}$ denote the numbers of antennas along the horizontal and vertical dimensions, respectively.
The $L$ satellites are interconnected via \acp{isl}, which enables cooperative transmission. We assume that the \acp{isl} can exchange the coordination information required by the proposed design, e.g., over a connected backbone, without committing to a particular routing or topology.
The serving set consists of $L$ satellites close to the centroid of the \acp{ut}, since the \ac{leo} channel is dominated by its \ac{los} component~\cite{you2020jsac,zack2025twc}, whose path loss grows with the satellite-\ac{ut} distance. The design takes the serving set as given, so any selection rule can be used, e.g., the one in Section~\ref{subsec:setup}.

Due to the high orbital velocity of \ac{leo} satellites (about $7.6~\text{km/s}$ at an altitude of $590~\text{km}$), the serving set varies over time. In contrast, the set of \acp{ut} evolves on a much slower timescale and is therefore assumed fixed. To account for the orbital dynamics in a tractable manner, we adopt a frame-wise model with $K$ frames, in which the serving set is fixed within each frame and updated at its beginning. The index set of the $L$ satellites selected in the $k$-th frame is $\mathcal{L}^{(k)}$.

\subsection{Downlink Signal Model\label{signal_first}}
In the $k$-th frame, the satellites in $\mathcal{L}^{(k)}$ cooperatively transmit $U$ independent data streams, collected in $\mathbf{s}^{(k)}=[s_1^{(k)},\ldots,s_U^{(k)}]^{\mathsf{T}}$ with $\mathbb{E}[\mathbf{s}^{(k)}\mathbf{s}^{(k)\mathsf{H}}]=\mathbf{I}_U$. We assume that these satellites are synchronized in time/frequency/phase and have access to the data streams of the \acp{ut} they serve (e.g., via feeder links or \acp{isl}~\cite{yafei2026jsac}) for coherent joint transmission~\cite{moewin2025jsac,zack2025twc}.
\begin{figure*}[t]
		\centering
		\begin{tikzpicture}[x=1cm, y=1cm, font=\footnotesize,
    bar/.style={draw=black!65, line width=0.4pt},
    data/.style={bar, fill=green!50!black!14},
    decision/.style={bar, fill=violet!13},
    note/.style={font=\scriptsize, inner sep=1pt},
    rowlab/.style={anchor=east, align=right, font=\footnotesize, inner sep=0pt},
    brace/.style={decorate, decoration={brace, amplitude=3pt}, line width=0.4pt}]
  \def\xo{3.30}\def\xa{8.75}\def\xb{14.20}\def\xc{16.85}
  \def\dt{2.15}
  \def\hb{0.36}\def\hf{0.40}
  \def\yF{3.22}\def\yA{2.54}\def\yB{1.86}\def\yC{1.18}
  \foreach \s in {\xo,\xa,\xb} \draw[decision] (\s,\yF) rectangle ++(\dt,\hf) node[midway, note] {Handover decision};
  \foreach \s in {\xo,\xa} \draw[bar, fill=white] (\s+\dt,\yF) rectangle (\s+5.45,\yF+\hf) node[midway, note] {Data transmission};
  \draw[bar] (\xb+\dt,\yF) -- (\xc,\yF) (\xb+\dt,\yF+\hf) -- (\xc,\yF+\hf);
  \node[rowlab] at (3.10,\yF+\hf/2) {Frame structure};
  \draw[brace] (\xo,\yF+0.47) -- node[above=2pt, note] {$T$} (\xa,\yF+0.47);
  \draw[brace] (\xa,\yF+0.47) -- node[above=2pt, note] {$\tau_{\mathrm{HO}}T$} (\xa+\dt,\yF+0.47);
  \draw[brace] (\xa+\dt,\yF+0.47) -- node[above=2pt, note] {$(1-\tau_{\mathrm{HO}})T$} (\xb,\yF+0.47);
  \draw[densely dotted, black!35, line width=0.4pt] (\xa+\dt,\yF) -- (\xa+\dt,\yC-0.06);
  \foreach \s in {\xa,\xb} \draw[densely dotted, black!50, line width=0.4pt] (\s,\yF) -- (\s,0.42);
  \draw[data] (\xo,\yA) rectangle (\xa,\yA+\hb) node[midway, note] {Data with $\mathbf{w}_{\ell,u}^{(k-1)}$};
  \draw[data] (\xa,\yA) rectangle (\xb,\yA+\hb) node[midway, note] {Data with $\mathbf{w}_{\ell,u}^{(k)}$ in both segments};
  \draw[data] (\xb,\yA) -- (\xc,\yA) (\xb,\yA+\hb) -- (\xc,\yA+\hb);
  \node[rowlab] at (3.10,\yA+\hb/2) {Persisting link\\[-1pt]{\scriptsize$\delta_{\ell,u}^{(k-1)}=\delta_{\ell,u}^{(k)}=1$}};
  \draw[data] (\xo,\yB) rectangle (\xa,\yB+\hb) node[midway, note] {Data with $\mathbf{w}_{\ell,u}^{(k-1)}$};
  \node[note, text=black!60, anchor=west] at (\xa+0.08,\yB+\hb/2) {Released at the frame boundary, $\mathbf{w}_{\ell,u}^{(k)}=\mathbf{0}$};
  \node[rowlab] at (3.10,\yB+\hb/2) {Released link\\[-1pt]{\scriptsize$\delta_{\ell,u}^{(k-1)}=1$, $\delta_{\ell,u}^{(k)}=0$}};
  \draw[decision] (\xa,\yC) rectangle ++(\dt,\hb) node[midway, note] {Setup, $E_{\mathrm{HO},\ell,u}^{(k)}$};
  \draw[data] (\xa+\dt,\yC) rectangle (\xb,\yC+\hb) node[midway, note] {Data with $\mathbf{w}_{\ell,u}^{(k)}$};
  \draw[data] (\xb,\yC) -- (\xc,\yC) (\xb,\yC+\hb) -- (\xc,\yC+\hb);
  \node[rowlab] at (3.10,\yC+\hb/2) {New link\\[-1pt]{\scriptsize$\delta_{\ell,u}^{(k-1)}=0$, $\delta_{\ell,u}^{(k)}=1$}};
  \draw[brace] (\xa+\dt,\yC-0.08) -- node[below=2pt, note, xshift=-0.22cm] {Rate $\tilde{R}_u^{(k)}$ (persisting links)} (\xa,\yC-0.08);
  \draw[brace] (\xb,\yC-0.08) -- node[below=2pt, note, xshift=0.12cm] {Rate $R_u^{(k)}$ (all links)} (\xa+\dt,\yC-0.08);
  \draw[-{Stealth[length=4pt]}, line width=0.4pt] (\xo,0.42) -- (\xc+1.35,0.42) node[anchor=south east, note, xshift=-3pt, yshift=1pt] {Time};
  \foreach \s in {\xo,\xa,\xb} \draw[line width=0.4pt] (\s,0.36) -- (\s,0.48);
  \node[note, anchor=north] at ({(\xo+\xa)/2},0.36) {Frame $k-1$};
  \node[note, anchor=north] at ({(\xa+\xb)/2},0.36) {Frame $k$};
  \node[note, anchor=north west] at (\xb,0.36) {Frame $k+1$};
  \foreach \y in {\yF+\hf/2,\yA+\hb/2,\yC+\hb/2} \node[note] at (\xc+0.35,\y) {$\cdots$};
\end{tikzpicture}%
		\caption{Frame structure and link types. Each frame of duration $T$ starts with a handover-decision subframe of duration $\tau_{\mathrm{HO}}T$. A persisting link carries data during the whole frame, a new link only after its setup, which consumes the handover energy $E_{\mathrm{HO},\ell,u}^{(k)}$, and a released link stops at the frame boundary. Hence, the $u$-th \ac{ut} obtains the rate $\tilde{R}_u^{(k)}$ in the handover-decision subframe, in which only its persisting links carry data, and $R_u^{(k)}$ in the data segment, in which all its links carry data.}
		\label{frame_struct}
\end{figure*}
The transmit signal of the $\ell$-th satellite is expressed as
\begin{equation}\label{eq_tx}
\mathbf{x}_{\ell}^{(k)}
= \sum_{u=1}^{U}\mathbf{w}_{\ell,u}^{(k)} s_u^{(k)}
= \mathbf{W}_{\ell}^{(k)} \mathbf{s}^{(k)},
\end{equation}
where $\mathbf{W}_{\ell}^{(k)}=[\mathbf{w}_{\ell,1}^{(k)},\ldots,\mathbf{w}_{\ell,U}^{(k)}]\in\mathbb{C}^{N\times U}$ stacks the beamforming vectors from the $\ell$-th satellite to all \acp{ut}. Since each satellite has a limited number of \acp{rfc}, it can simultaneously serve at most $U_{\rm{max}}$ \acp{ut}, i.e., at most $U_{\rm{max}}$ columns of $\mathbf{W}_{\ell}^{(k)}$ can be nonzero, so that user scheduling arises for $U_{\max}<U$.

In the $k$-th frame, the received signal at the $u$-th \ac{ut} is
\begin{align}\label{eq_rx}
y_u^{(k)}
&= \sum_{\ell \in\mathcal{L}^{(k)}} \mathbf{h}_{\ell,u}^{(k)\mathsf{T}} \mathbf{x}_{\ell}^{(k)} + n_u^{(k)} \\*
&= \sum_{\ell \in\mathcal{L}^{(k)}} \mathbf{h}_{\ell,u}^{(k)\mathsf{T}} \mathbf{w}_{\ell,u}^{(k)} s_u^{(k)}
\!+\!\underbrace{\sum_{i\neq u}^{U}\sum_{\ell \in\mathcal{L}^{(k)}} \mathbf{h}_{\ell,u}^{(k)\mathsf{T}} \mathbf{w}_{\ell,i}^{(k)} s_i^{(k)}}_{\text{\Acf{iui}}}
\!+\! n_u^{(k)}, \notag
\end{align}
where $\mathbf{h}_{\ell,u}^{(k)}\in\mathbb{C}^{N}$ denotes the channel vector from the $\ell$-th satellite to the $u$-th \ac{ut} and $n_u^{(k)} \sim \mathcal{CN}(0,\sigma^2)$ denotes the \ac{awgn}, with variance $\sigma^2 = FN_0 B$. Here, $F$ is the noise figure, $N_0$ the single-sided \ac{psd}, and $B$ the system bandwidth.

\subsection{LEO Channel Representation\label{subsec_channel}}
We next specify $\mathbf{h}_{\ell,u}^{(k)}$, starting from a multipath model and simplifying it in two steps. In the $k$-th frame, the multipath model between the $\ell$-th satellite and the $u$-th \ac{ut} at time $t$ and frequency $f$ is expressed by
\begin{align}\label{eq_chan_raw}
\mathbf{h}_{\ell,u}^{(k)}(t,f)
= &\sum_{m=0}^{M_{\ell,u}^{(k)}} \alpha_{\ell,u,m}^{(k)}
G\left(\boldsymbol{\theta}^{(k)}_{\ell,u,m}\right)
e^{\jmath 2\pi\left(t\upsilon_{\ell,u,m}^{(k)}-f\tau_{\ell,u,m}^{(k)}\right)}\notag \\*
&\times \mathbf{a}\left(\boldsymbol{\theta}_{\ell,u,m}^{(k)}\right),
\end{align}
where $M_{\ell,u}^{(k)}$ denotes the number of \ac{nlos} paths, indexed by $m\ge 1$, while $m=0$ denotes the \ac{los} path, and $\alpha_{\ell,u,m}^{(k)}$ represents the complex channel gain associated with the $m$-th path. Here, $\tau_{\ell,u,m}^{(k)}$ and $\upsilon_{\ell,u,m}^{(k)}$ denote the delay and the Doppler shift of the $m$-th path,
$\boldsymbol{\theta}_{\ell,u,m}^{(k)}=[\theta_{\ell,u,m}^{(k)\rm{az}},\theta_{\ell,u,m}^{(k)\rm{el}}]^{\mathsf{T}}$
is its \ac{aod}, and $G(\boldsymbol{\theta})$ is the amplitude radiation pattern of an array element in direction $\boldsymbol{\theta}$.
Let $\mathbf{n}(N)=[0,\ldots,N-1]^{\mathsf{T}}$.
The steering vector is
\begin{equation}\label{steer }
\mathbf{a}\left(\boldsymbol{\theta}_{\ell,u,m}^{(k)}\right)
=
e^{-\jmath 2\pi \phi^{(k)\rm{h}}_{\ell,u,m}\mathbf{n}(N_{\rm{h}})}
\otimes
e^{-\jmath 2\pi \phi^{(k)\rm{v}}_{\ell,u,m}\mathbf{n}(N_{\rm{v}})},
\end{equation}
where, for half-wavelength spacing, the spatial frequencies are
\begin{align}\label{eq_spatial_freq}
\phi^{(k)\rm{h}}_{\ell,u,m}&=\frac{1}{2}\cos\theta^{(k)\rm{az}}_{\ell,u,m}\cos\theta^{(k)\rm{el}}_{\ell,u,m},\notag\\*
\phi^{(k)\rm{v}}_{\ell,u,m}&=\frac{1}{2}\sin\theta^{(k)\rm{az}}_{\ell,u,m}\cos\theta^{(k)\rm{el}}_{\ell,u,m}.
\end{align}

\subsubsection{LoS Approximation of LEO Channel}
Since \ac{leo} satellites are far above the local scattering region around each \ac{ut}, all paths of a link leave the satellite in nearly the same direction, i.e.,
$\boldsymbol{\theta}_{\ell,u,m}^{(k)}\approx \boldsymbol{\theta}_{\ell,u}^{(k)}$ for all $m$.
We also decompose the Doppler shift into satellite- and \ac{ut}-induced components as
$\upsilon_{\ell,u,m}^{(k)}=\upsilon_{\ell,u,m}^{(k)\rm{Sat}}+\upsilon_{\ell,u,m}^{(k)\rm{UT}}$.
The satellite-induced component is much larger than the \ac{ut}-induced one and nearly the same for all paths, i.e.,
$\upsilon_{\ell,u,m}^{(k)\rm{Sat}}\approx \upsilon_{\ell,u}^{(k)\rm{Sat}}$ for all $m$.

Let $\tau_{\ell,u}^{(k)} = \tau_{\ell,u,0}^{(k)}$ and $\upsilon_{\ell,u}^{(k)} = \upsilon_{\ell,u,0}^{(k)}$ denote the delay and the Doppler shift of the \ac{los} path, and define the differential delays and Doppler shifts
\begin{align}\label{eq_diff}
\tau^{(k)\rm{Diff}}_{\ell,u,m}&= \tau_{\ell,u,m}^{(k)}-\tau_{\ell,u}^{(k)},\notag\\*
\upsilon^{(k)\rm{Diff}}_{\ell,u,m}&= \upsilon_{\ell,u,m}^{(k)}-\upsilon_{\ell,u}^{(k)}\approx\upsilon_{\ell,u,m}^{(k)\rm{UT}}-\upsilon_{\ell,u,0}^{(k)\rm{UT}}.
\end{align}
With the above approximations, \eqref{eq_chan_raw} can be rearranged into
\begin{equation}\label{eq_chan_rank1}
\mathbf{h}_{\ell,u}^{(k)} \left(t,f\right)
=
\alpha_{\ell,u}^{(k)} \left(t,f\right)\,
e^{\jmath 2\pi\left(t\upsilon_{\ell,u}^{(k)}-f\tau_{\ell,u}^{(k)}\right)}
\mathbf{b} \left(\boldsymbol{\theta}_{\ell,u}^{(k)}\right),
\end{equation}
where $\mathbf{b}(\boldsymbol{\theta}_{\ell,u}^{(k)})= G(\boldsymbol{\theta}^{(k)}_{\ell,u})\mathbf{a}(\boldsymbol{\theta}_{\ell,u}^{(k)})$
combines the array response and the element gain, and the effective channel gain is
\begin{equation}\label{eq_alpha_def}
\alpha_{\ell,u}^{(k)} \left(t,f\right)
=
\sum_{m=0}^{M_{\ell,u}^{(k)}}\alpha_{\ell,u,m}^{(k)}
e^{\jmath 2\pi\left(t\upsilon_{\ell,u,m}^{(k)\rm{Diff}}-f\tau_{\ell,u,m}^{(k)\rm{Diff}}\right)}.
\end{equation}
The \ac{los} term of \eqref{eq_alpha_def}, whose differential delay and Doppler shift are zero, is constant, whereas the \ac{nlos} terms make $\alpha_{\ell,u}^{(k)}(t,f)$ fluctuate over $t$ and $f$. Since the statistics of these fluctuations depend only on the geometry, which changes little within a frame (Section~\ref{subsec:setup}), we take them as constant over the frame and the bandwidth. Hence, we drop the $(t,f)$ dependence and adopt a standard Rician model with Rician factor $\kappa_{\ell,u}^{(k)}$ and average power $\mathbb{E} [|\alpha_{\ell,u}^{(k)}|^2]=\gamma_{\ell,u}^{(k)}$~\cite{you2020jsac}. Specifically, $\alpha_{\ell,u}^{(k)}=\bar{\alpha}_{\ell,u}^{(k)}+\tilde{\alpha}_{\ell,u}^{(k)}$, where $\tilde{\alpha}_{\ell,u}^{(k)}\sim\mathcal{CN}(0,\beta_{\ell,u}^{(k)})$ is the scattered part, and the \ac{los} amplitude $\bar{\alpha}_{\ell,u}^{(k)}$ and the variance $\beta_{\ell,u}^{(k)}$ are
\begin{equation}\label{eq_rician_params}
\bar{\alpha}_{\ell,u}^{(k)}=\sqrt{\frac{\kappa_{\ell,u}^{(k)}\gamma_{\ell,u}^{(k)}}{1+\kappa_{\ell,u}^{(k)}}},
\qquad
\beta_{\ell,u}^{(k)}=\frac{\gamma_{\ell,u}^{(k)}}{1+\kappa_{\ell,u}^{(k)}},
\end{equation}
so that $\bar{\alpha}_{\ell,u}^{(k)2}+\beta_{\ell,u}^{(k)}=\gamma_{\ell,u}^{(k)}$. The deterministic phase of the \ac{los} component can be absorbed into $\mathbf{b}(\boldsymbol{\theta}_{\ell,u}^{(k)})$ in \eqref{eq_chan_rank1}, so $\bar{\alpha}_{\ell,u}^{(k)}$ is taken real and nonnegative without loss of generality.

\subsubsection{Ephemeris-Aided Compensation}
The satellite position and velocity are known from the ephemeris~\cite[Sec.~16.14.2.2]{3gpp.38.300}, so the dominant delay and Doppler terms in \eqref{eq_chan_rank1} can be compensated, e.g., at each \ac{ut}~\cite{you2020jsac,kexin2024twc}. For coherent joint transmission, however, a \ac{ut} receives the sum of signals with different delays and Doppler shifts, which a single correction at the \ac{ut} cannot align. As in~\cite{yafei2026jsac}, we therefore assume that the delays and Doppler shifts are \emph{pre-compensated at the transmitter} per data stream, i.e., each satellite compensates the \ac{los} delay and Doppler shift in \eqref{eq_chan_rank1}, obtained from the ephemeris and the \ac{ut} position and velocity~\cite{zack2025twc,moewin2025jsac}, on the stream intended for each \ac{ut} before radiation. These delays and Doppler shifts also differ across \acp{ut}, so such per-stream pre-compensation aligns each stream only at its intended \ac{ut} and may induce asynchronous interference terms in a strict signal model (see, e.g.,~\cite{asynLEO2024TWC,yafei2026jsac}). However, since this paper focuses on the handover-aware design, we adopt the standard abstraction of ideal compensation and leave these asynchrony effects to future work. Under this assumption, the effective channel used in \eqref{eq_rx} reduces to
\begin{equation}\label{eq_chan_final}
\mathbf{h}_{\ell,u}^{(k)}=\alpha_{\ell,u}^{(k)}\mathbf{b}\!\left(\boldsymbol{\theta}_{\ell,u}^{(k)}\right).
\end{equation}

\section{Handover-Aware Power Consumption and Communication Performance Metrics\label{sec_handover}}
In this section, we introduce a two-segment frame structure that separates the handover procedure from the user-plane data transmission, and then define the corresponding handover-aware power consumption model and statistical rate metric, on which the joint design in Section~\ref{sec_method} is based.

\subsection{Frame Structure}
Because of the orbital motion, the \ac{leo}-\ac{ut} schedule is inherently time-varying. To preserve link visibility and quality, each \ac{ut} must periodically switch its serving satellites. Each new link requires a handover procedure, i.e., signaling, synchronization, and link establishment, before user-plane data can be transmitted over it~\cite{3gpp.38.300, juan2022access}.
For example, in \ac{3gpp} \ac{nr}-\ac{ntn}, when a handover to a new serving satellite/beam is triggered, the \ac{ut} pre-compensates its uplink timing advance and Doppler shift using its position and the satellite ephemeris~\cite[Sec.~16.14.2.2]{3gpp.38.300}, and typically performs a random access channel (RACH) procedure in the target cell~\cite[Sec.~9.2.3]{3gpp.38.300}.
This procedure consumes dedicated physical random access channel (PRACH) resources and additional transmission power for control-plane signaling.
In a networked \ac{leo} system, we assume that the same procedure is performed whenever a new satellite-\ac{ut} link is established, while the other cooperating satellites of the \ac{ut} keep transmitting.

Motivated by this procedure, we adopt the frame structure in Fig.~\ref{frame_struct}, in which each frame is partitioned into two segments. The first segment, called the handover-decision subframe, occupies a fraction $\tau_{\mathrm{HO}}\in(0,1)$ of the frame. During this subframe, the handovers decided for the frame are executed, i.e., the new links of the frame are established. In the remaining fraction $1-\tau_{\mathrm{HO}}$, called the data segment, all links of the current schedule carry user-plane data. Links that persist from the previous frame carry data in both segments.

\subsection{Handover-Aware Power Consumption Model}
Let $\delta_{\ell,u}^{(k)} = 1_{+}\{\|\mathbf{w}_{\ell,u}^{(k)}\|^2\}$ indicate whether the $\ell$-th satellite serves the $u$-th \ac{ut} in the $k$-th frame, i.e., $\delta_{\ell,u}^{(k)}=1$ if and only if the corresponding beamformer is nonzero. By comparing $\delta_{\ell,u}^{(k-1)}$ with $\delta_{\ell,u}^{(k)}$, we distinguish three types of links in the frame structure of Fig.~\ref{frame_struct}. A link that persists from the $(k-1)$-th to the $k$-th frame, i.e., $\delta_{\ell,u}^{(k-1)}=\delta_{\ell,u}^{(k)}=1$, requires no handover and carries data during the whole frame with the current beamformer $\mathbf{w}_{\ell,u}^{(k)}$. A new link, i.e., $\delta_{\ell,u}^{(k-1)}=0$ and $\delta_{\ell,u}^{(k)}=1$, carries data only after the handover-decision subframe, during which its access signaling and synchronization are performed on dedicated control resources. A released link, i.e., $\delta_{\ell,u}^{(k-1)}=1$ and $\delta_{\ell,u}^{(k)}=0$, stops at the frame boundary. A satellite $\ell\in\mathcal{L}^{(k)}\setminus\mathcal{L}^{(k-1)}$ that newly enters the serving set has no established links, i.e., $\delta_{\ell,u}^{(k-1)}=0$ for all $u$.

Accordingly, the power consumption of the $\ell$-th satellite in the $k$-th frame is modeled as
\begin{align}\label{per_power_model}
P_{\ell}^{(k)}
=& \sum_{u=1}^{U}\Bigg(
\underbrace{\delta_{\ell,u}^{(k)}\big(1-\delta_{\ell,u}^{(k-1)}\big) P_{\mathrm{HO},\ell,u}^{(k)}}_{\text{Handover-related power}}
\\*
&
+ \underbrace{\tau_{\mathrm{HO}}\big\|\delta_{\ell,u}^{(k-1)}\mathbf{w}_{\ell,u}^{(k)}\big\|^2
+ (1-\tau_{\mathrm{HO}})\big\|\mathbf{w}_{\ell,u}^{(k)}\big\|^2}_{\text{Radiated power}}
\Bigg),\notag
\end{align}
where the handover-related term is nonzero only for a new link, and the radiated-power term counts a persisting link in both segments, a new link only in the second, and a released link, for which $\mathbf{w}_{\ell,u}^{(k)}=\mathbf{0}$, in neither. Here, $P_{\mathrm{HO},\ell,u}^{(k)}=E_{\mathrm{HO},\ell,u}^{(k)}/T$ is the energy $E_{\mathrm{HO},\ell,u}^{(k)}$ spent to establish a new link between the $\ell$-th satellite and the $u$-th \ac{ut} (e.g., for access, synchronization, and control signaling~\cite{3gpp.38.300}), averaged over the frame duration $T$, so that every term of \eqref{per_power_model} is a power. As this signaling must reach the \ac{ut} at a given signal-to-noise ratio, its energy grows with the free-space loss, and we model
\begin{equation}\label{eq_pho}
P_{\mathrm{HO},\ell,u}^{(k)}=P_{\mathrm{HO}}\left(\frac{d_{\ell,u}^{(k)}}{h}\right)^{2},
\end{equation}
where $d_{\ell,u}^{(k)}$ is the slant range, $h$ is the orbit altitude, and $P_{\mathrm{HO}}=E_{\mathrm{HO}}/T$ is the handover-related power of a link to a \ac{ut} at the sub-satellite point, i.e., with $d_{\ell,u}^{(k)}=h$. Although this overhead is typically smaller than the radiated power, it is nonzero and can accumulate under frequent handovers. We therefore model it explicitly to capture its impact on the network power budget. As its value depends on the access procedure and the signaling load, Section~\ref{subsec:ho} varies $P_{\mathrm{HO}}$ over two orders of magnitude.

The network-wide power consumption in the $k$-th frame is
\begin{equation}\label{total_power_model}
P_{\mathrm{total}}^{(k)} = \sum_{\ell \in\mathcal{L}^{(k)}} P_{\ell}^{(k)},
\end{equation}
which depends on the previous-frame schedule through $\delta_{\ell,u}^{(k-1)}$.

\begin{remark}[Make-Before-Break Handover through Cooperation]\label{rem:continuity}
A \ac{ut} with a persisting link keeps its service while its new links are established, since this link carries data during the whole frame, i.e., its handover is make-before-break. This relies on multi-connectivity, i.e., on a \ac{ut} connected to several satellites at once, which the cooperative transmission of Section~\ref{sysmod} assumes. Multi-connectivity is studied for \acp{ntn} in~\cite{majamaa2024mcom}, and dual connectivity was used in~\cite{lema2026tcom} for make-before-break handover at the \ac{leo} \ac{ut}. Unlike dual active protocol stack (DAPS) handover, in which the \ac{ut} keeps receiving data from the source cell until releasing it~\cite[Sec.~9.2.3.1]{3gpp.38.300} and which \ac{nr}-\ac{ntn} does not support~\cite[Sec.~16.14.3.2.1]{3gpp.38.300}, the proposed model keeps no released link during the handover. A \ac{ut} whose links are all new, e.g., a handed-over \ac{ut} with a single link, receives data only after the handover-decision subframe, i.e., its handover is break-before-make.
\end{remark}

\subsection{Handover-Aware Communication Performance Metric}
In \ac{leo} systems, accurate instantaneous \ac{csi} is hard to acquire, due to the rapid channel variation induced by the satellite motion and the long round-trip delay~\cite{kexin2023twc,chang2023iotj,semiblind2025cl,blockKF2023cl,ming2025twc}. We therefore evaluate the communication performance with a statistical metric. Specifically, we adopt the hardening bound, a widely used lower bound on the ergodic rate that depends only on the channel statistics~\cite{Marzetta2016mMIMO,caire2018twc}. In the $k$-th frame, let
\begin{equation}\label{eq_g}
g_{\ell,u,i}^{(k)}= \mathbf{b}^{\mathsf{T}}\left(\boldsymbol{\theta}_{\ell,u}^{(k)}\right)\mathbf{w}_{\ell,i}^{(k)}
\end{equation}
denote the beamformed gain of the $i$-th stream at the $u$-th \ac{ut} through the array of the $\ell$-th satellite, and let $\mathbf{g}_{u,i}^{(k)},\boldsymbol{\alpha}_u^{(k)}\in\mathbb{C}^{L}$ stack $g_{\ell,u,i}^{(k)}$ and $\alpha_{\ell,u}^{(k)}$ over $\ell\in\mathcal{L}^{(k)}$ in the same order. With \eqref{eq_chan_final}, the received signal \eqref{eq_rx} becomes
\begin{align}
y_u^{(k)}&=\sum_{i=1}^{U}\Upsilon_{u,i}^{(k)}s_i^{(k)}+n_u^{(k)},\label{eq_rx_agg}\\*
\Upsilon_{u,i}^{(k)} &= \sum_{\ell\in\mathcal{L}^{(k)}} \alpha_{\ell,u}^{(k)} g_{\ell,u,i}^{(k)},\label{eq_Upsilon_def}
\end{align}
where $\Upsilon_{u,i}^{(k)}$ is the aggregated gain of the $i$-th stream at the $u$-th \ac{ut}. The hardening bound takes the mean $\mathbb{E}[\Upsilon_{u,u}^{(k)}]$ as the desired-signal gain and treats its deviation from this mean and the interference as uncorrelated noise, which yields a lower bound on the ergodic rate of the $u$-th \ac{ut}~\cite{Marzetta2016mMIMO,zack2025twc,zack2025decentral}
\begin{equation}\label{eq_rate_lb}
R_u^{(k)}
\!=\!
\log_2 \!\left(\!
1\!+\!\frac{\big|\mathbb{E}\left[\Upsilon_{u,u}^{(k)}\right]\big|^2}{
\mathbb{V}\left[\Upsilon_{u,u}^{(k)}\right] \!+\! \sum_{i\neq u}^{U}\mathbb{E}\left[|\Upsilon_{u,i}^{(k)}|^2\right]\! + \!\sigma^2}
\right).
\end{equation}

By using the Rician parameterization in \eqref{eq_rician_params} and assuming independence of $\{\alpha_{\ell,u}^{(k)}\}_{\ell}$ across satellites (which is reasonable since the satellites are far apart and seen from the \ac{ut} in different directions), we obtain
\begin{subequations}\label{eq_rate_parts}
\begin{align}
\mathbb{E} \left[\Upsilon_{u,u}^{(k)}\right]
&=\sum_{\ell \in\mathcal{L}^{(k)}}\bar{\alpha}_{\ell,u}^{(k)}g_{\ell,u,u}^{(k)},\\*
\mathbb{V} \left[\Upsilon_{u,u}^{(k)}\right]
&=\sum_{\ell \in\mathcal{L}^{(k)}}\beta_{\ell,u}^{(k)}\left|g_{\ell,u,u}^{(k)}\right|^2,\\*
\mathbb{E} \left[\left|\Upsilon_{u,i}^{(k)}\right|^2\right]
&=\mathbf{g}_{u,i}^{(k)\mathsf{H}}\mathbf{T}_u^{(k)}\mathbf{g}_{u,i}^{(k)}.
\end{align}
\end{subequations}
Here, the correlation matrix of $\boldsymbol{\alpha}_u^{(k)}$ is
\begin{equation}\label{eq_Tu}
\mathbf{T}_u^{(k)} = \mathbb{E} \left[\boldsymbol{\alpha}_u^{(k)}\boldsymbol{\alpha}_u^{(k)\mathsf{H}}\right]
=
\bar{\boldsymbol{\alpha}}_u^{(k)}\bar{\boldsymbol{\alpha}}_u^{(k)\mathsf{H}}+\rm{diag}(\boldsymbol{\beta}_u^{(k)}),
\end{equation}
where $\bar{\boldsymbol{\alpha}}_u^{(k)}$ and $\boldsymbol{\beta}_u^{(k)}$ stack $\bar{\alpha}_{\ell,u}^{(k)}$ and $\beta_{\ell,u}^{(k)}$ over $\ell\in\mathcal{L}^{(k)}$, so that $R_u^{(k)}$ depends on the channel only through these statistical parameters and the steering vectors.

Introducing the handover at the beginning of each frame leads to a segmented performance metric. Since new links carry data only after the handover-decision subframe, we consider the time-weighted average of the rates of the two segments:
\begin{equation}
R_{u}^{(k),\mathrm{seg}} = \tau_{\mathrm{HO}} \tilde{R}_u^{(k)} + \left(1-\tau_{\mathrm{HO}}\right) R_u^{(k)},
\end{equation}
where $R_u^{(k)}$ is the rate of the data segment, in which all links carry data, and $\tilde{R}_u^{(k)}$ is the rate of the handover-decision subframe, in which only the persisting links carry data. The rate $\tilde{R}_u^{(k)}$ follows from $R_u^{(k)}$ by replacing $g_{\ell,u,i}^{(k)}$ in \eqref{eq_rate_parts} with
\begin{equation}\label{eq_g_tilde}
\tilde{g}_{\ell,u,i}^{(k)} = \mathbf{b}^{\mathsf{T}}\left(\boldsymbol{\theta}_{\ell,u}^{(k)}\right)\delta_{\ell,i}^{(k-1)}\mathbf{w}_{\ell,i}^{(k)}.
\end{equation}
The indicator in \eqref{eq_g_tilde} is indexed by the stream $i$, not by the receiving \ac{ut} $u$, since during the handover-decision subframe satellite $\ell$ radiates stream $i$ only if the link $(\ell,i)$ persists from the previous frame, consistently with the radiated-power term in \eqref{per_power_model}. This stream also interferes at the \acp{ut} $u\neq i$.

\section{Handover-Aware Joint Cooperative Beamforming and User Scheduling\label{sec_method}}
Based on the power model and the rate metric of Section~\ref{sec_handover}, this section formulates a per-frame network-wide power minimization problem and develops an iterative algorithm that (i) handles the sparsity induced by the scheduling via a reweighted $\ell_2$ approximation, (ii) converts the \ac{qos} constraint of the first frame into an equivalent convex form, and (iii) handles those of the subsequent frames by an \ac{fp}-based inner loop.

\subsection{Problem Formulation}
For the $k$-th frame, we minimize the network-wide power consumption, including the handover cost, by jointly optimizing cooperative beamforming and \ac{leo}-\ac{ut} scheduling:
\begin{subequations}\label{ori_prob}
\begin{align}
\min_{\{\mathbf{w}_{\ell,u}^{(k)}\}_{\forall \ell,u}} \;
& P_{\mathrm{total}}^{(k)} \label{ori_prob_obj}\\*
\mathrm{s.t.}\;
& \tau_{\mathrm{HO}} \tilde{R}_u^{(k)} + \left(1-\tau_{\mathrm{HO}}\right) R_u^{(k)}
\ge \bar{R}_{u}, \; \forall u, \label{qos_constraint}\\*
& \sum_{u=1}^{U} 1_{+}\{\|\mathbf{w}_{\ell,u}^{(k)}\|^2\} \le U_{\mathrm{max}}, \; \forall \ell, \label{rfc_constraint}\\*
& \sum_{u=1}^{U} \left\|\mathbf{w}_{\ell,u}^{(k)}\right\|^2 \le P_{\mathrm{rad}}, \; \forall \ell. \label{sig_pow_constraint}
\end{align}
\end{subequations}
Here, \eqref{qos_constraint} enforces the \ac{qos} requirement for each \ac{ut} with a minimum rate of $\bar{R}_{u}$, \eqref{rfc_constraint} limits each satellite to serving at most $U_{\mathrm{max}}$ \acp{ut}, thereby inducing user scheduling, and \eqref{sig_pow_constraint} limits the radiated power of each satellite to $P_{\mathrm{rad}}$. Although the beamformers $\{\mathbf{w}_{\ell,u}^{(k)}\}_{\forall \ell,u}$ are the explicit optimization variables, the schedule is implicitly determined by which beamformers are nonzero, as captured by $1_{+}\{\|\mathbf{w}_{\ell,u}^{(k)}\|^2\}$, so no binary scheduling variables are needed.

Problem \eqref{ori_prob} is challenging to solve for two reasons. First, user scheduling is combinatorial due to the indicators in \eqref{rfc_constraint} and in the handover-related term of \eqref{per_power_model}. Second, the \ac{qos} constraints \eqref{qos_constraint} are nonconvex in the beamformers. The coupling between frames adds no further difficulty, since the previous-frame schedule $\delta_{\ell,u}^{(k-1)}$ is fixed when the $k$-th frame is designed and thus enters \eqref{ori_prob} only as a parameter of \eqref{per_power_model} and \eqref{eq_g_tilde}. Sections~\ref{sec_method}-B and~\ref{sec_method}-C handle the two difficulties.

\subsection{Transformation of Sparsity Constraint}
The serving-cardinality constraint \eqref{rfc_constraint} is nonconvex because it enforces sparsity on $\{\mathbf{w}_{\ell,u}^{(k)}\}_{\forall \ell, u}$. To obtain a convex constraint, we express this sparsity through auxiliary power weights $\omega_{\ell,u}^{(k)}$, one for each beamformer, and apply the iterative reweighting of~\cite{boyd2008sparsity} to their squares, i.e., a reweighted $\ell_2$-norm approximation. Specifically, we impose
\begin{equation}\label{per_bf_power}
\left\|\mathbf{w}_{\ell,u}^{(k)}\right\| \le \omega_{\ell,u}^{(k)} \sqrt{P_{\mathrm{rad}}}, \; \forall \ell,u.
\end{equation}
Since \eqref{sig_pow_constraint} implies $\|\mathbf{w}_{\ell,u}^{(k)}\|\le\sqrt{P_{\mathrm{rad}}}$, the weights can be restricted without loss of generality to
\begin{equation}\label{power_weight_constraint}
0 \le \omega_{\ell,u}^{(k)} \le 1, \; \forall \ell,u.
\end{equation}
By \eqref{per_bf_power}, $\omega_{\ell,u}^{(k)2}$ is an upper bound on the fraction $\|\mathbf{w}_{\ell,u}^{(k)}\|^2/P_{\mathrm{rad}}$ of the power budget of the $\ell$-th satellite allocated to the $u$-th \ac{ut}. We then apply the sparsity indicator to the weights, i.e., we replace
$1_{+}\{\|\mathbf{w}_{\ell,u}^{(k)}\|^2\}$ with $1_{+}\{\omega_{\ell,u}^{(k)}\}$, which yields
\begin{equation}\label{sparsity_new}
\sum_{u=1}^{U} 1_{+}\left\{\omega_{\ell,u}^{(k)}\right\} \le U_{\mathrm{max}}, \; \forall \ell.
\end{equation}
This replacement is exact, since $\omega_{\ell,u}^{(k)}=0$ forces $\mathbf{w}_{\ell,u}^{(k)}=\mathbf{0}$ by \eqref{per_bf_power}, and $\omega_{\ell,u}^{(k)}=\|\mathbf{w}_{\ell,u}^{(k)}\|/\sqrt{P_{\mathrm{rad}}}$ satisfies \eqref{per_bf_power} and \eqref{power_weight_constraint}. To handle the remaining nonconvexity in \eqref{sparsity_new}, we approximate
$1_{+}\{\omega_{\ell,u}^{(k)}\}$ with a weighted quadratic surrogate, whose reweighting coefficients $z_{\ell,u}^{(k)}$ are updated from the previous iterate. Given the beamformers $\tilde{\mathbf{w}}_{\ell,u}^{(k)}$ of the previous iteration and $\tilde{\omega}_{\ell,u}^{(k)}=\|\tilde{\mathbf{w}}_{\ell,u}^{(k)}\|/\sqrt{P_{\mathrm{rad}}}$, we set
\begin{equation}\label{l2_weight}
z_{\ell,u}^{(k)} = \frac{1}{\tilde{\omega}_{\ell,u}^{(k)2}+\epsilon}, \; \forall \ell,u,
\end{equation}
where $\epsilon>0$ avoids a division by zero and is set relative to the largest weight (Table~\ref{tab:simu_para}). Then we employ the approximation
\begin{equation}\label{l2_approx}
1_{+}\left\{\omega_{\ell,u}^{(k)}\right\} \approx z_{\ell,u}^{(k)} \omega_{\ell,u}^{(k)2}, \; \forall \ell,u,
\end{equation}
which leads to a convex surrogate of \eqref{sparsity_new} as
\begin{equation}\label{rfc_constraint_re}
\sum_{u=1}^{U} z_{\ell,u}^{(k)} \omega_{\ell,u}^{(k)2} \le U_{\mathrm{max}}, \; \forall \ell.
\end{equation}
At $\omega_{\ell,u}^{(k)}=\tilde{\omega}_{\ell,u}^{(k)}$, the surrogate \eqref{l2_approx} equals $\tilde{\omega}_{\ell,u}^{(k)2}/(\tilde{\omega}_{\ell,u}^{(k)2}+\epsilon)$, which is close to one for a link with $\tilde{\omega}_{\ell,u}^{(k)2}\gg\epsilon$ and zero for an inactive link. Moreover, since $z_{\ell,u}^{(k)}$ is large for links with small previous weights, the reweighting pushes these weights toward zero and promotes sparsity.

Enforcing \eqref{rfc_constraint_re} at every iteration can make a subproblem infeasible, since links with small previous weights obtain large coefficients $z_{\ell,u}^{(k)}$, and \eqref{rfc_constraint_re} then allows them only small weights $\omega_{\ell,u}^{(k)}$, even if the \ac{qos} requires larger ones. We therefore relax \eqref{rfc_constraint_re} by a slack variable $\zeta_{\ell}^{(k)}$ that is penalized in the objective~\cite{gui2020tsp,jorge2006numerical}. In addition, the handover-related term of $P_{\mathrm{total}}^{(k)}$ contains the indicator $\delta_{\ell,u}^{(k)}=1_{+}\{\|\mathbf{w}_{\ell,u}^{(k)}\|^2\}$, which we replace by the same surrogate \eqref{l2_approx}, while $\delta_{\ell,u}^{(k-1)}$ is fixed by the previous frame. This yields the convex objective
\begin{align}\label{eq_P_hat}
\hat{P}_{\mathrm{total}}^{(k)}=\!\!\sum_{\ell\in\mathcal{L}^{(k)}}\sum_{u=1}^{U}\!\Big(&\big(1-\delta_{\ell,u}^{(k-1)}\big)P_{\mathrm{HO},\ell,u}^{(k)}z_{\ell,u}^{(k)}\omega_{\ell,u}^{(k)2}\notag\\*
&\hspace{-1.6cm}+\tau_{\mathrm{HO}}\big\|\delta_{\ell,u}^{(k-1)}\mathbf{w}_{\ell,u}^{(k)}\big\|^2+(1-\tau_{\mathrm{HO}})\big\|\mathbf{w}_{\ell,u}^{(k)}\big\|^2\Big).
\end{align}%
Problem \eqref{ori_prob} is thus approximated by
\begin{subequations}\label{rfc_constraint_relax}
\begin{align}
&\min_{\{\mathbf{w}_{\ell,u}^{(k)},\omega_{\ell,u}^{(k)},\zeta_{\ell}^{(k)}\}_{\forall \ell,u}} \;
 \hat{P}_{\mathrm{total}}^{(k)} + \mu \sum\nolimits_{\ell\in\mathcal{L}^{(k)}}\zeta_{\ell}^{(k)} \\*
\mathrm{s.t.}\;
&\tau_{\mathrm{HO}} \tilde{R}_u^{(k)} + \left(1-\tau_{\mathrm{HO}}\right) R_u^{(k)}
\ge \bar{R}_{u}, \; \forall u,\\*
&\sum_{u=1}^{U} z_{\ell,u}^{(k)} \omega_{\ell,u}^{(k)2} - U_{\mathrm{max}} \le \zeta_{\ell}^{(k)}, \; \zeta_{\ell}^{(k)}\ge 0, \; \forall \ell,\\*
&\left\|\mathbf{w}_{\ell,u}^{(k)}\right\| \le \omega_{\ell,u}^{(k)} \sqrt{P_{\mathrm{rad}}}, \; \omega_{\ell,u}^{(k)}\le 1, \; \forall \ell,u,\\*
&\sum_{u=1}^{U} \left\|\mathbf{w}_{\ell,u}^{(k)}\right\|^2 \le P_{\mathrm{rad}}, \; \forall \ell,
\end{align}
\end{subequations}
where $\mu$ is a penalty coefficient that controls how strongly violations of \eqref{rfc_constraint_re} are discouraged. Following~\cite{gui2020tsp}, $\mu$ is increased across iterations as $\mu \leftarrow \min\{\rho \mu,\mu_{\mathrm{max}}\}$ with a constant $\rho>1$, where the upper bound $\mu_{\mathrm{max}}$ avoids numerical issues. When $\mu$ becomes large, $\zeta_{\ell}^{(k)}$ is driven close to zero, so that \eqref{rfc_constraint_re} is approximately enforced. The only nonconvex constraint left in \eqref{rfc_constraint_relax} is the \ac{qos} constraint (\ref{rfc_constraint_relax}b), which we transform next.

\subsection{Transformation of QoS Constraint}

\subsubsection{Case \texorpdfstring{$k=1$}{k=1}}
No links exist before the first frame, i.e., $\delta_{\ell,u}^{(0)}=0$, so no data is transmitted during its handover-decision subframe and $\tilde{R}_u^{(1)}=0$. Hence, \eqref{qos_constraint} becomes
\begin{equation}\label{f1_qos_constraint}
R_u^{(1)} \ge \frac{\bar{R}_{u}}{1-\tau_{\mathrm{HO}}}, \; \forall u.
\end{equation}

Constraint \eqref{f1_qos_constraint} is nonconvex, since $R_u^{(1)}$ contains a quadratic-over-quadratic \ac{sinr} term, but it admits an equivalent \ac{soc} representation, as we show next. Define
\begin{equation}\label{eq_eta_f1}
\eta_u = 2^{\bar{R}_u/(1-\tau_{\mathrm{HO}})}-1,
\end{equation}
and
$\mathbf{g}_{u}^{(1)} = [\mathbf{g}_{u,1}^{(1)\mathsf{T}},\ldots,\mathbf{g}_{u,U}^{(1)\mathsf{T}}]^{\mathsf{T}}$.
Then \eqref{f1_qos_constraint} is equivalent to
\begin{gather}
\frac{\left| \sum_{\ell \in \mathcal{L}^{(1)}} \bar{\alpha}_{\ell,u}^{(1)} g_{\ell,u,u}^{(1)} \right|^2}
{\mathbf{g}_{u}^{(1)\mathsf{H}}\boldsymbol{\Omega}_u^{(1)}\mathbf{g}_{u}^{(1)} + \sigma^2}
\ge \eta_u,\; \forall u,\label{sinr_constraint}\\
\text{with}~\boldsymbol{\Omega}_u^{(1)} = \mathrm{blkdiag}\left(\mathbf{T}_u^{(1)},\ldots,\underbrace{\mathbf{Q}_u^{(1)}}_{\text{$u$-th block}},\ldots,\mathbf{T}_u^{(1)}\right),\label{eq_Omega_def}
\end{gather}
and $\mathbf{Q}_u^{(1)} = \mathrm{diag}\big([\beta_{\ell,u}^{(1)}]_{\forall \ell\in\mathcal{L}^{(1)}}\big)$. The $u$-th block of $\boldsymbol{\Omega}_u^{(1)}$ yields the variance (\ref{eq_rate_parts}b) and the others the interference (\ref{eq_rate_parts}c).

Moreover, a \emph{common} phase rotation of $\{\mathbf{w}_{\ell,u}^{(1)}\}_{\forall \ell\in\mathcal{L}^{(1)}}$ (and hence of $\{g_{\ell,u,u}^{(1)}\}_{\forall \ell\in\mathcal{L}^{(1)}}$) changes neither the objective nor the constraints. Hence, without loss of optimality, we impose
\begin{subequations}\label{rotation_constraint}
\begin{align}
&\Re \left\{\sum_{\ell \in\mathcal{L}^{(1)}}\bar{\alpha}_{\ell,u}^{(1)}g_{\ell,u,u}^{(1)}\right\} \ge 0, \; \forall u, \label{real_constraint}\\*
&\Im \left\{\sum_{\ell \in\mathcal{L}^{(1)}}\bar{\alpha}_{\ell,u}^{(1)}g_{\ell,u,u}^{(1)}\right\} = 0, \; \forall u.   \label{imag_constraint}
\end{align}
\end{subequations}
With \eqref{rotation_constraint}, \eqref{sinr_constraint} can be equivalently written in \ac{soc} form as
\begin{equation}\label{eq_soc_form}
\left\|\big[(\boldsymbol{\Psi}_u^{(1)}\mathbf{g}_{u}^{(1)})^{\mathsf{T}},\ \sigma\big]^{\mathsf{T}}\right\|
\le \frac{1}{\sqrt{\eta_u}} \sum_{\ell \in \mathcal{L}^{(1)}} \bar{\alpha}_{\ell,u}^{(1)} g_{\ell,u,u}^{(1)}, \; \forall u,
\end{equation}
where $\boldsymbol{\Psi}_u^{(1)}$ satisfies
$\boldsymbol{\Omega}_u^{(1)}=\boldsymbol{\Psi}_u^{(1)\mathsf{H}}\boldsymbol{\Psi}_u^{(1)}$
(e.g., a Cholesky factor), which exists since $\boldsymbol{\Omega}_u^{(1)}$ is positive semidefinite. Constraint \eqref{real_constraint} is implied by \eqref{eq_soc_form} and can be dropped, whereas \eqref{imag_constraint} keeps the right-hand side of \eqref{eq_soc_form} real.

\subsubsection{Case \texorpdfstring{$k>1$}{k>1}}
When $k>1$, we first decouple the weighted two-segment \ac{qos} constraint in \eqref{qos_constraint} by introducing auxiliary variables $\tilde{\Gamma}_u^{(k)}$ and $\Gamma_u^{(k)}$, yielding
\begin{align}
&\tau_{\mathrm{HO}} \tilde{\Gamma}_u^{(k)} + \left(1-\tau_{\mathrm{HO}}\right) \Gamma_u^{(k)}
\ge \bar{R}_{u}, \; \forall u,\label{segment_qos_constraint_p1}  \\*
&\tilde{R}_u^{(k)} \ge \tilde{\Gamma}_u^{(k)}, \; \forall u,\label{segment_qos_constraint_p2}\\*
&R_u^{(k)} \ge \Gamma_u^{(k)}, \; \forall u.\label{segment_qos_constraint_p3}
\end{align}
It remains to handle the nonconvex constraints \eqref{segment_qos_constraint_p2} and \eqref{segment_qos_constraint_p3}.

For a \ac{ut} whose links are all new, i.e., $\delta_{\ell,u}^{(k-1)}=0$ for all $\ell$, we have $\tilde{R}_u^{(k)}=0$, so its constraint reduces to the single-rate form \eqref{f1_qos_constraint} and is handled exactly by \eqref{eq_soc_form}, as for $k=1$. For the other \acp{ut}, \eqref{segment_qos_constraint_p2} and \eqref{segment_qos_constraint_p3} are equivalent to
\begin{align}
&\frac{\big|\sum_{\ell \in \mathcal{L}^{(k)}} \bar{\alpha}_{\ell,u}^{(k)} \tilde{g}_{\ell,u,u}^{(k)} \big|^2}
{\tilde{\mathbf{g}}_{u}^{(k)\mathsf{H}}\boldsymbol{\Omega}_u^{(k)}\tilde{\mathbf{g}}_{u}^{(k)} + \sigma^2}  \ge 2^{\tilde{\Gamma}_u^{(k)}}  - 1, \label{segment_rotation_p1_a} \\
&\frac{\big| \sum_{\ell \in \mathcal{L}^{(k)}} \bar{\alpha}_{\ell,u}^{(k)} g_{\ell,u,u}^{(k)} \big|^2}
{\mathbf{g}_{u}^{(k)\mathsf{H}}\boldsymbol{\Omega}_u^{(k)}\mathbf{g}_{u}^{(k)} + \sigma^2}  \ge 2^{\Gamma_u^{(k)}}  - 1,\label{segment_rotation_p2_a}
\end{align}
respectively, where $\tilde{\mathbf{g}}_{u}^{(k)}$ and $\mathbf{g}_{u}^{(k)}$ stack $\tilde{\mathbf{g}}_{u,i}^{(k)}$ and $\mathbf{g}_{u,i}^{(k)}$ over $i$ as for $k=1$, and $\boldsymbol{\Omega}_u^{(k)}$ is defined as in \eqref{eq_Omega_def}. Unlike the case $k=1$, a common phase rotation of $\{\mathbf{w}_{\ell,u}^{(k)}\}_{\ell}$ cannot make the sums in the numerators of \eqref{segment_rotation_p1_a} and \eqref{segment_rotation_p2_a} real at the same time, because the sum in \eqref{segment_rotation_p1_a} contains only the previously established links. Hence, the \ac{soc} form \eqref{eq_soc_form} does not apply, and we handle the fractions in \eqref{segment_rotation_p1_a} and \eqref{segment_rotation_p2_a} by \ac{fp} instead, with complex auxiliary variables that need no phase constraint.

Specifically, using the quadratic transform of~\cite{kaiming2018tsp,kaiming2025spm}, also applied in~\cite{zack2024tsp}, we transform \eqref{segment_rotation_p1_a} and \eqref{segment_rotation_p2_a} into
\begin{align}
\mathop {\max }\limits_{\tilde{\lambda}_u^{(k)}\in\mathbb{C}}\; &2\Re\Big\{ \tilde{\lambda}_u^{(k)*}\sum_{\ell \in \mathcal{L}^{(k)}} \bar{\alpha}_{\ell,u}^{(k)} \tilde{g}_{\ell,u,u}^{(k)}\Big\} \notag \\*
&\hspace{-0.3cm}- \big|\tilde{\lambda}_u^{(k)}\big|^2\big(\tilde{\mathbf{g}}_{u}^{(k)\mathsf{H}}\boldsymbol{\Omega}_u^{(k)}\tilde{\mathbf{g}}_{u}^{(k)} + \sigma^2\big) \ge 2^{\tilde{\Gamma}_u^{(k)}}  - 1,\label{segment_fp_p1}\\
\mathop {\max }\limits_{\lambda_u^{(k)}\in\mathbb{C}}\; &2\Re\Big\{ \lambda_u^{(k)*}\sum_{\ell \in \mathcal{L}^{(k)}} \bar{\alpha}_{\ell,u}^{(k)} g_{\ell,u,u}^{(k)}\Big\} \notag \\*
&\hspace{-0.3cm}- \big|\lambda_u^{(k)}\big|^2\big(\mathbf{g}_{u}^{(k)\mathsf{H}}\boldsymbol{\Omega}_u^{(k)}\mathbf{g}_{u}^{(k)} + \sigma^2\big) \ge 2^{\Gamma_u^{(k)}}  - 1,\label{segment_fp_p2}
\end{align}
respectively, where $\tilde{\lambda}_u^{(k)}$ and $\lambda_u^{(k)}$ are the introduced auxiliary variables. For any $a\in\mathbb{C}$, $b>0$, and $\lambda\in\mathbb{C}$, it holds that
\begin{equation*}
\frac{|a|^2}{b}-\left(2\Re\{\lambda^{*}a\}-|\lambda|^2 b\right)=\frac{|a-\lambda b|^2}{b}\ge 0,
\end{equation*}
with equality if and only if $\lambda=a/b$. For fixed auxiliary variables, the left-hand sides of \eqref{segment_fp_p1} and \eqref{segment_fp_p2} are concave in the beamformers and lower-bound the \acp{sinr}, with equality when
\begin{align}
\tilde{\lambda}_u^{(k)} &= \frac{\sum_{\ell \in \mathcal{L}^{(k)}} \bar{\alpha}_{\ell,u}^{(k)} \tilde{g}_{\ell,u,u}^{(k)}}
{\tilde{\mathbf{g}}_{u}^{(k)\mathsf{H}}\boldsymbol{\Omega}_u^{(k)}\tilde{\mathbf{g}}_{u}^{(k)} + \sigma^2},\label{fp_aux_p1}\\
\lambda_u^{(k)} &= \frac{\sum_{\ell \in \mathcal{L}^{(k)}} \bar{\alpha}_{\ell,u}^{(k)} g_{\ell,u,u}^{(k)}}
{\mathbf{g}_{u}^{(k)\mathsf{H}}\boldsymbol{\Omega}_u^{(k)}\mathbf{g}_{u}^{(k)} + \sigma^2}.\label{fp_aux_p2}
\end{align}
Therefore, we alternate between solving the problem with fixed auxiliary variables and updating them by \eqref{fp_aux_p1} and \eqref{fp_aux_p2}, which makes the bounds tight at the current beamformers.

\renewcommand{\algorithmicrequire}{\textbf{Input:}}
\renewcommand{\algorithmicensure}{\textbf{Output:}}
\begin{algorithm}[t]
\caption{Proposed Handover-Aware Joint Cooperative Beamforming and User Scheduling}
\label{proposed_algo}
\begin{algorithmic}[1]
\Require $\{\bar{\alpha}_{\ell,u}^{(k)},\beta_{\ell,u}^{(k)},\mathbf{b}(\boldsymbol{\theta}_{\ell,u}^{(k)}),\mathcal{L}^{(k)}\}_{\forall \ell,u,k}$, $\tau_{\mathrm{HO}}$, $U_{\max}$, $P_{\mathrm{rad}}$, $P_{\mathrm{HO}}$, $\sigma^2$, $\epsilon$, $\rho$, $\mu_{\max}$, $N_{\mathrm{rw}}$, $N_{\mathrm{fp}}$, and $\mu=\mu_0$.
\State \textbf{Initialize}: $\{\mathbf{w}_{\ell,u}^{(1)},\tilde{\omega}_{\ell,u}^{(1)},z_{\ell,u}^{(1)}\}_{\forall \ell,u}$.
\For{$k=1:K$}
    \If{$k=1$ (\emph{Initial frame})}
        \Repeat { (\emph{Loop: reweighted $\ell_2$ approximation})}
            \State Solve \eqref{f1_prob_re}.
            \State Update $z_{\ell,u}^{(1)}$ using \eqref{l2_weight}.
            \State Update $\mu \leftarrow \min\{\rho \mu,\mu_{\rm{max}}\}$.
        \Until{the set of active links is unchanged and satisfies \eqref{rfc_constraint}, or $N_{\mathrm{rw}}$ iterations are reached.}
        \State Extract the schedule and refine $\{\mathbf{w}_{\ell,u}^{(1)}\}$.
        \State Set $\delta_{\ell,u}^{(1)} = 1_{+}\{\|\mathbf{w}_{\ell,u}^{(1)}\|^2\}$.
    \Else {~(\emph{Subsequent frames})}
        \State Initialize $\{\mathbf{w}_{\ell,u}^{(k)},\tilde{\omega}_{\ell,u}^{(k)},z_{\ell,u}^{(k)}\}_{\forall \ell,u}$ and set $\mu=\mu_0$.
        \Repeat {~(\emph{Outer loop: reweighted $\ell_2$ approximation})}
            \State Initialize $\{\tilde{\lambda}_u^{(k)},\lambda_u^{(k)}\}_{\forall u}$ via \eqref{fp_aux_p1} and \eqref{fp_aux_p2}.
            \Repeat {~(\emph{Inner loop: \ac{fp} process})}
                \State Solve \eqref{fk_prob_re}.
                \State Update $\{\tilde{\lambda}_u^{(k)},\lambda_u^{(k)}\}_{\forall u}$ via \eqref{fp_aux_p1} and \eqref{fp_aux_p2}.
            \Until{the relative change of the objective of \eqref{fk_prob_re} falls below a threshold, or $N_{\mathrm{fp}}$ iterations are reached.}
            \State Update $z_{\ell,u}^{(k)}$ using \eqref{l2_weight}.
            \State Update $\mu \leftarrow \min\{\rho \mu,\mu_{\rm{max}}\}$.
        \Until{the set of active links is unchanged and satisfies \eqref{rfc_constraint}, or $N_{\mathrm{rw}}$ iterations are reached.}
        \State Extract the schedule and refine $\{\mathbf{w}_{\ell,u}^{(k)}\}$.
        \State Set $\delta_{\ell,u}^{(k)} = 1_{+}\{\|\mathbf{w}_{\ell,u}^{(k)}\|^2\}$.
    \EndIf
\EndFor
\State \textbf{Output}: $\{\mathbf{w}_{\ell,u}^{(k)}\}$ and $\{\delta_{\ell,u}^{(k)}\}$.
\end{algorithmic}
\end{algorithm}

\subsection{Overall Procedure}
Algorithm~\ref{proposed_algo} summarizes the procedure for $k=1$ and $k>1$.

\subsubsection{Case \texorpdfstring{$k=1$}{k=1}}
At each iteration, given $z_{\ell,u}^{(1)}$, we replace (\ref{rfc_constraint_relax}b) with the convex constraints \eqref{imag_constraint} and \eqref{eq_soc_form}, and solve
\begin{align}\label{f1_prob_re}
&\min_{\{\mathbf{w}_{\ell,u}^{(1)},\omega_{\ell,u}^{(1)},\zeta_{\ell}^{(1)}\}_{\forall \ell,u}}\;\hat{P}_{\mathrm{total}}^{(1)} +\mu\!\!\sum_{\ell\in\mathcal{L}^{(1)}}\!\!\zeta_{\ell}^{(1)}\notag\\*
&\ \mathrm{s.t.}\ \eqref{imag_constraint},\ \eqref{eq_soc_form},\ \text{(\ref{rfc_constraint_relax}c)--(\ref{rfc_constraint_relax}e)}\text{ with }k=1.
\end{align}
Problem \eqref{f1_prob_re} is a \acf{socp}, which can be solved by a standard tool such as CVX. We then update $z_{\ell,u}^{(1)}$ from the resulting beamformers via \eqref{l2_weight}, increase $\mu$, and repeat until the stopping rule of Algorithm~\ref{proposed_algo} is met.

\begin{figure*}[t]
\centering
\includegraphics[width=1\linewidth]{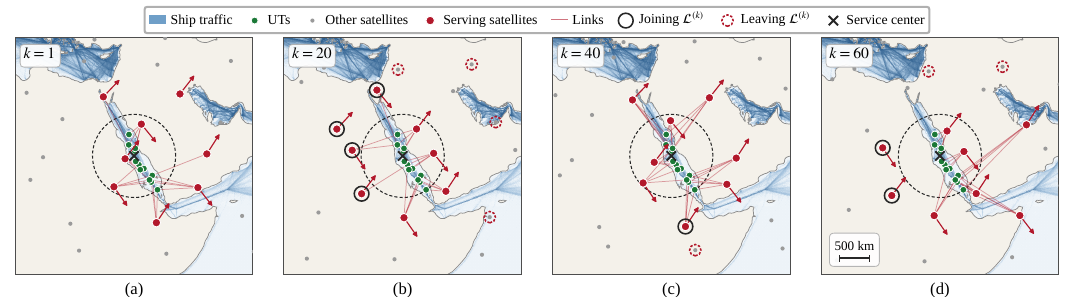}
\caption{Serving satellite set $\mathcal{L}^{(k)}$ and links of the proposed design in frames $k=1$, $20$, $40$, and $60$ of the trajectory studied in Section~\ref{subsec:time}. Satellites are drawn at their sub-satellite points, the arrows show their displacement during the next minute, and the lines connect each serving satellite with the \acp{ut} that it serves. A satellite is marked as joining or leaving if it joined or left the serving set after the frame of the previous panel. The background shows the ship traffic density of~\cite{worldbank2021shipdensity}, and the dashed circle the service area.}
\label{topology_snapshots}
\end{figure*}
\subsubsection{Case \texorpdfstring{$k>1$}{k>1}}
When $k>1$, the \ac{fp}-based reformulation introduces the auxiliary variables $\tilde{\lambda}_u^{(k)}$ and $\lambda_u^{(k)}$, which leads to a double-loop procedure. In each outer iteration, we fix $z_{\ell,u}^{(k)}$. Then, in each inner iteration, we also fix $\tilde{\lambda}_u^{(k)}$ and $\lambda_u^{(k)}$, and solve the convex subproblem
\begin{align}\label{fk_prob_re}
&\min_{\{\mathbf{w}_{\ell,u}^{(k)},\omega_{\ell,u}^{(k)},\tilde{\Gamma}_u^{(k)},\Gamma_u^{(k)},\zeta_{\ell}^{(k)}\}_{\forall \ell,u}}\;\hat{P}_{\mathrm{total}}^{(k)} + \mu\!\!\sum_{\ell\in\mathcal{L}^{(k)}}\!\!\zeta_{\ell}^{(k)}\notag\\*
&\ \mathrm{s.t.}\ \eqref{segment_qos_constraint_p1},\ \eqref{segment_fp_p1},\ \eqref{segment_fp_p2},\ \text{(\ref{rfc_constraint_relax}c)--(\ref{rfc_constraint_relax}e)},
\end{align}
where \eqref{segment_fp_p1} and \eqref{segment_fp_p2} are replaced by \eqref{imag_constraint} and \eqref{eq_soc_form} for \acp{ut} whose links are all new. Problem \eqref{fk_prob_re} is convex for fixed $z_{\ell,u}^{(k)}$, $\tilde{\lambda}_u^{(k)}$, and $\lambda_u^{(k)}$: the left-hand sides of \eqref{segment_fp_p1} and \eqref{segment_fp_p2} are then concave quadratic functions of the beamformers, their right-hand sides are convex functions of $\tilde{\Gamma}_u^{(k)}$ and $\Gamma_u^{(k)}$, and \eqref{segment_qos_constraint_p1} and (\ref{rfc_constraint_relax}c)--(\ref{rfc_constraint_relax}e) are convex constraints. Hence, \eqref{fk_prob_re} can be solved by standard convex solvers, e.g., CVX. After each inner iteration, the resulting beamformers are used to update $\tilde{\lambda}_u^{(k)}$ and $\lambda_u^{(k)}$ via \eqref{fp_aux_p1} and \eqref{fp_aux_p2}, and the inner loop repeats until its stopping rule in Algorithm~\ref{proposed_algo} is met. The outer loop then updates $z_{\ell,u}^{(k)}$ from the resulting beamformers via \eqref{l2_weight}, increases $\mu$, and repeats until its stopping rule is met.

\subsubsection{Initialization and Schedule Extraction\label{sec_readout}}
The \ac{fp} updates must start from a point that satisfies \eqref{qos_constraint}. In each frame, the iterations therefore start from the solution of \eqref{f1_prob_re}, written for the $k$-th frame, with $z_{\ell,u}^{(k)}=1$. In this problem, every \ac{ut} has the single-rate constraint \eqref{f1_qos_constraint}, which implies \eqref{qos_constraint} since $\tilde{R}_u^{(k)}\ge0$. If this problem is infeasible, such a point is found by minimizing the total shortfall of the rates below $\bar{R}_u$.

\leavevmode The reweighting and the penalty promote \eqref{rfc_constraint} but do not enforce it, so a satellite may still serve more than $U_{\max}$ \acp{ut} after the outer loop. We therefore extract a schedule with at most $U_{\max}$ links per satellite and at least one link per \ac{ut}, and refine its beamformers, i.e., re-optimize them by \ac{fp} iterations. For this, we form two candidate schedules from the link powers $\|\mathbf{w}_{\ell,u}^{(k)}\|^2$ of the starting point and two from those of every outer iterate. In the first, each satellite keeps its strongest links. In the second, it keeps its active links, i.e., those whose power exceeds $1\%$ of the largest link power, and adds its links of the previous frame, which need no handover. After the refinement, the candidate with the lowest $P_{\mathrm{total}}^{(k)}$ is kept.

\subsubsection{Convergence}
\leavevmode For fixed reweighting coefficients $z_{\ell,u}^{(k)}$ and penalty $\mu$, the current beamformers are feasible for \eqref{fk_prob_re} after the auxiliary variables are updated, since the bounds in \eqref{segment_fp_p1} and \eqref{segment_fp_p2} are tight at \eqref{fp_aux_p1} and \eqref{fp_aux_p2}. Hence, the objective of \eqref{fk_prob_re} does not increase over the inner iterations and, being bounded below, converges. The same holds for the refinement of each candidate schedule once a feasible starting point is found. The outer loop stops once the set of active links is the same as in the previous iteration and satisfies \eqref{rfc_constraint}, or after the maximum number $N_{\mathrm{rw}}$ of outer iterations. Hence, Algorithm~\ref{proposed_algo} terminates in every frame, and its beamformers satisfy all constraints of \eqref{ori_prob} whenever the refinement is feasible.

\subsubsection{Complexity Analysis}
Each subproblem, i.e., \eqref{f1_prob_re} for $k=1$ and \eqref{fk_prob_re} for $k>1$, is a convex problem solved by an interior-point method, whose worst-case complexity scales as $\mathcal{O}(n^{3.5})$ in the number $n$ of variables. The dominant variables are the beamformers $\{\mathbf{w}_{\ell,u}^{(k)}\}_{\forall \ell,u}$, whose total dimension is $|\mathcal{L}^{(k)}|UN$. Since $\mathbf{w}_{\ell,i}^{(k)}$ enters the problem only through $\{\mathbf{b}^{\mathsf{T}}(\boldsymbol{\theta}_{\ell,u}^{(k)})\mathbf{w}_{\ell,i}^{(k)}\}_{u}$ and $\|\mathbf{w}_{\ell,i}^{(k)}\|$, it can be restricted without loss of optimality to the span of $\{\mathbf{b}^{*}(\boldsymbol{\theta}_{\ell,u}^{(k)})\}_{u}$, so $N$ can be replaced by $\min\{N,U\}$ in the following orders. Therefore, the per-frame complexity of the outer loop scales as $\mathcal{O}(N_{\mathrm{rw}}(|\mathcal{L}^{(1)}|UN)^{3.5})$ for $k=1$, and as $\mathcal{O}(N_{\mathrm{rw}}N_{\mathrm{fp}}(|\mathcal{L}^{(k)}|UN)^{3.5})$ for $k>1$, where $N_{\mathrm{rw}}$ and $N_{\mathrm{fp}}$ denote the maximum numbers of outer reweighting iterations and inner \ac{fp} iterations, respectively. In addition, the schedule extraction refines at most $2(N_{\mathrm{rw}}+1)$ candidate schedules, i.e., two for the starting point and two for each outer iterate. Each refinement runs \ac{fp} iterations on a convex problem of the same form, which contains only the beamformers of the candidate links, i.e., at most $|\mathcal{L}^{(k)}|U_{\max}$ beamformers instead of $|\mathcal{L}^{(k)}|U$.

\section{Numerical Results\label{sec_num}}
In this section, we present simulation results to evaluate the proposed design over trajectories of a networked \ac{leo} system and to compare it with benchmark schemes.

\subsection{Simulation Setup\label{subsec:setup}}
We consider a Walker-Delta constellation, a common model of \ac{leo} mega-constellations~\cite{xiaoming2025twc,yafei2026jsac}, with $28$ orbit planes of $28$ satellites at an altitude of $590~\mathrm{km}$, an inclination of $53^\circ$, and phasing factor $1$, whose circular orbits are propagated around a spherical Earth with radius $6371~\mathrm{km}$. A trajectory consists of $K$ frames of duration $T=5~\mathrm{s}$. Within a frame, a satellite moves by about $38~\mathrm{km}$, which is small compared with its slant range of at least $590~\mathrm{km}$, so we evaluate the statistical \ac{csi} of each frame at its middle. The \acp{ut} are vessels in the central Red Sea, a busy shipping lane. Like the maritime traffic model of~\cite{eva2020mari}, we use ship positions reported by the \ac{ais}: in each Monte-Carlo trial, we draw the $U$ vessels from the \ac{ais}-based ship traffic density of~\cite{worldbank2021shipdensity,cerdeiro2020imf}, within $700$~km of a service center at latitude $20^\circ$ and longitude $38.5^\circ$ and at least $20$~km off the coast, i.e., away from ports (see Fig.~\ref{topology_snapshots}), and treat them as static. Their motion is negligible: even at $15~\mathrm{m/s}$, i.e., about $30$ knots, a vessel moves by less than $5~\mathrm{km}$ during the longest trajectory, of $5$~min, whereas a satellite moves by about $2300~\mathrm{km}$. The serving set of the first frame consists of the $L$ satellites closest to the centroid of the \acp{ut}. Afterwards, a member is replaced only by a satellite that is closer to the centroid by more than $10\%$ of the member's distance, which keeps satellites from leaving and re-joining the set within a few frames.

For each frame, the average power $\gamma_{\ell,u}^{(k)}$ of the link between the $\ell$-th satellite and the $u$-th \ac{ut} is the inverse of its large-scale path loss, which is generated according to~\cite{3gpp.38.811,zack2025twc}. The Rician factor $\kappa_{\ell,u}^{(k)}$ is drawn uniformly from $[15,20]$~dB, within the $10$--$20$~dB range noted for the $12$~GHz Ku band in~\cite{moewin2025jsac}. The shadow fading and the Rician factor of each link are drawn once per trajectory, so that the statistics change continuously with the geometry. The element pattern follows~\cite[Table~\mbox{7.3-1}]{3gpp.38.901} for a nadir-facing panel, with a maximum gain of $8$~dBi, a $3$-dB beamwidth of $65^\circ$, and a maximum attenuation of $30$~dB.

\looseness=-1 Unless stated otherwise, the parameters are those of Table~\ref{tab:simu_para}, and the refinement of the extracted schedule stops once its objective changes by less than $0.1\%$, or after $10$ \ac{fp} iterations. In each Monte-Carlo trial, the start time of the trajectory is drawn uniformly within one day, and all schemes are evaluated on the same vessels, trajectories, shadow fading, and Rician factors. A frame in which a scheme cannot satisfy all constraints is infeasible for this scheme: its power is excluded from the averages, and all its links are released, so that every link of the next frame is new. We therefore report the feasibility rate, i.e., the fraction of feasible frames, and the power saving of the proposed design over each other scheme on the frames in which both are feasible, if these are at least $4\%$ of all frames. A negative saving is the saving of the other scheme over the proposed design.

\begin{table}[t]
    \centering
    \caption{Default simulation parameters}
    \label{tab:simu_para}
    \setlength{\tabcolsep}{4pt}
    \begin{tabular}{@{}ll@{}}
        \toprule
        \textbf{Parameter} & \textbf{Value} \\
        \midrule
        Carrier frequency $f_c$ & $12~\mathrm{GHz}$, Ku band~\cite{todd2023taes} \\
        System bandwidth $B$ & $250~\mathrm{MHz}$, channel spacing~\cite{todd2023taes} \\
        Noise \ac{psd} $N_0$ & $-173.855~\mathrm{dBm/Hz}$ \\
        Noise figure $F$ & $4~\mathrm{dB}$ \\
        Number of satellites in the serving set $L$ & $8$ \\
        Number of \acp{ut} $U$ & $12$ \\
        UPA size $N = N_{\rm{h}} \times N_{\rm{v}}$ & $8\times 8$ \\
        Serving cardinality $U_{\max}$ & $4$ \\
        Per-satellite power budget $P_{\mathrm{rad}}$ & $60~\mathrm{dBm}$ \\
        Handover-related power $P_{\mathrm{HO}}$ & $40~\mathrm{dBm}$, i.e., $E_{\mathrm{HO}}=50~\mathrm{J}$ \\
        Handover fraction $\tau_{\mathrm{HO}}$  & $0.1$ \\
        Minimum rate requirement $\bar{R}_{u}$ & $0.1~\mathrm{bps/Hz}$, i.e., $25~\mathrm{Mbps}$ \\
        Frame duration $T$, number of frames $K$ & $5~\mathrm{s}$, $24$ \\
        Reweighting constant $\epsilon$ & $10^{-2}\max_{\ell,u}\tilde{\omega}_{\ell,u}^{(k)2}$ \\
        Penalty $\mu_0$, $\rho$, $\mu_{\max}$ & $0.05P_{\mathrm{rad}}$, $2$, $50P_{\mathrm{rad}}$ \\
        Iterations $N_{\mathrm{rw}}$, $N_{\mathrm{fp}}$ & $30$, $1$ \\
        Monte-Carlo trials & $50$ \\
        \bottomrule
    \end{tabular}
\end{table}

\begin{figure}[t]
\centering
\begin{minipage}[b]{0.98\linewidth}
\vspace{-0.3cm}
  \centering
    \definecolor{cblue}{rgb}{0.066,0.443,0.745}%
\definecolor{corange}{rgb}{0.866,0.329,0.000}%
\definecolor{cyellow}{rgb}{0.929,0.694,0.125}%
\definecolor{cpurple}{rgb}{0.521,0.086,0.819}%
\definecolor{cgreen}{rgb}{0.231,0.666,0.196}%
\definecolor{ccyan}{rgb}{0.184,0.745,0.937}%
\definecolor{cmagenta}{rgb}{0.819,0.015,0.545}%
\begin{tikzpicture}
\begin{axis}[%
width=2.83in, height=0.62in, scale only axis, unbounded coords=jump,
xmin=1, xmax=60, ymin=0.3, ymax=15.7,
xlabel style={font=\footnotesize}, xlabel={Frame index $k$},
ylabel style={font=\footnotesize}, ylabel={Satellite (by entry)},
axis background/.style={fill=white}, xmajorgrids, ymajorgrids,
legend style={legend cell align=left, align=left, draw=white!15!black, font=\scriptsize, inner sep=1pt, row sep=-1.5pt, draw=none}, xtick={1,10,20,...,60}, ytick={1,15}, ytick style={draw=none}
]
\draw[black!45, densely dotted, line width=0.5pt] (axis cs:7,0.3) -- (axis cs:7,15.7);
\draw[black!45, densely dotted, line width=0.5pt] (axis cs:11,0.3) -- (axis cs:11,15.7);
\draw[black!45, densely dotted, line width=0.5pt] (axis cs:19,0.3) -- (axis cs:19,15.7);
\draw[black!45, densely dotted, line width=0.5pt] (axis cs:20,0.3) -- (axis cs:20,15.7);
\draw[black!45, densely dotted, line width=0.5pt] (axis cs:36,0.3) -- (axis cs:36,15.7);
\draw[black!45, densely dotted, line width=0.5pt] (axis cs:51,0.3) -- (axis cs:51,15.7);
\draw[black!45, densely dotted, line width=0.5pt] (axis cs:54,0.3) -- (axis cs:54,15.7);
\addplot [color=cblue, line width=2.2pt, forget plot]
  table[row sep=crcr]{%
1 1\\
2 1\\
3 1\\
4 1\\
5 1\\
6 1\\
7 1\\
8 1\\
9 1\\
10 1\\
11 1\\
12 1\\
13 1\\
14 1\\
15 1\\
16 1\\
17 1\\
18 1\\
19 1\\
20 1\\
21 1\\
22 1\\
23 1\\
24 1\\
25 1\\
26 1\\
27 1\\
28 1\\
29 1\\
30 1\\
31 1\\
32 1\\
33 1\\
34 1\\
35 1\\
36 1\\
37 1\\
38 1\\
39 1\\
40 1\\
41 1\\
42 1\\
43 1\\
44 1\\
45 1\\
46 1\\
47 1\\
48 1\\
49 1\\
50 1\\
51 1\\
52 1\\
53 1\\
54 1\\
55 1\\
56 1\\
57 1\\
58 1\\
59 1\\
60 1\\
};
\addplot [color=corange, line width=2.2pt, forget plot]
  table[row sep=crcr]{%
1 2\\
2 2\\
3 2\\
4 2\\
5 2\\
6 2\\
7 2\\
8 2\\
9 2\\
10 2\\
11 2\\
12 2\\
13 2\\
14 2\\
15 2\\
16 2\\
17 2\\
18 2\\
19 2\\
};
\addplot [color=cyellow, line width=2.2pt, forget plot]
  table[row sep=crcr]{%
1 3\\
2 3\\
3 3\\
4 3\\
5 3\\
6 3\\
7 3\\
8 3\\
9 3\\
10 3\\
11 3\\
12 3\\
13 3\\
14 3\\
15 3\\
16 3\\
17 3\\
18 3\\
19 3\\
20 3\\
21 3\\
22 3\\
23 3\\
24 3\\
25 3\\
26 3\\
27 3\\
28 3\\
29 3\\
30 3\\
31 3\\
32 3\\
33 3\\
34 3\\
35 3\\
36 3\\
37 3\\
38 3\\
39 3\\
40 3\\
41 3\\
42 3\\
43 3\\
44 3\\
45 3\\
46 3\\
47 3\\
48 3\\
49 3\\
50 3\\
};
\addplot [color=cpurple, line width=2.2pt, forget plot]
  table[row sep=crcr]{%
1 4\\
2 4\\
3 4\\
4 4\\
5 4\\
6 4\\
};
\addplot [color=cgreen, line width=2.2pt, forget plot]
  table[row sep=crcr]{%
1 5\\
2 5\\
3 5\\
4 5\\
5 5\\
6 5\\
7 5\\
8 5\\
9 5\\
10 5\\
};
\addplot [color=ccyan, line width=2.2pt, forget plot]
  table[row sep=crcr]{%
1 6\\
2 6\\
3 6\\
4 6\\
5 6\\
6 6\\
7 6\\
8 6\\
9 6\\
10 6\\
11 6\\
12 6\\
13 6\\
14 6\\
15 6\\
16 6\\
17 6\\
18 6\\
19 6\\
20 6\\
21 6\\
22 6\\
23 6\\
24 6\\
25 6\\
26 6\\
27 6\\
28 6\\
29 6\\
30 6\\
31 6\\
32 6\\
33 6\\
34 6\\
35 6\\
36 6\\
37 6\\
38 6\\
39 6\\
40 6\\
41 6\\
42 6\\
43 6\\
44 6\\
45 6\\
46 6\\
47 6\\
48 6\\
49 6\\
50 6\\
51 6\\
52 6\\
53 6\\
54 6\\
55 6\\
56 6\\
57 6\\
58 6\\
59 6\\
60 6\\
};
\addplot [color=cmagenta, line width=2.2pt, forget plot]
  table[row sep=crcr]{%
1 7\\
2 7\\
3 7\\
4 7\\
5 7\\
6 7\\
7 7\\
8 7\\
9 7\\
10 7\\
11 7\\
12 7\\
13 7\\
14 7\\
15 7\\
16 7\\
17 7\\
18 7\\
};
\addplot [color=cblue, line width=2.2pt, forget plot]
  table[row sep=crcr]{%
1 8\\
2 8\\
3 8\\
4 8\\
5 8\\
6 8\\
7 8\\
8 8\\
9 8\\
10 8\\
11 8\\
12 8\\
13 8\\
14 8\\
15 8\\
16 8\\
17 8\\
18 8\\
19 8\\
20 8\\
21 8\\
22 8\\
23 8\\
24 8\\
25 8\\
26 8\\
27 8\\
28 8\\
29 8\\
30 8\\
31 8\\
32 8\\
33 8\\
34 8\\
35 8\\
};
\addplot [color=corange, line width=2.2pt, forget plot]
  table[row sep=crcr]{%
7 9\\
8 9\\
9 9\\
10 9\\
11 9\\
12 9\\
13 9\\
14 9\\
15 9\\
16 9\\
17 9\\
18 9\\
19 9\\
20 9\\
21 9\\
22 9\\
23 9\\
24 9\\
25 9\\
26 9\\
27 9\\
28 9\\
29 9\\
30 9\\
31 9\\
32 9\\
33 9\\
34 9\\
35 9\\
36 9\\
37 9\\
38 9\\
39 9\\
40 9\\
41 9\\
42 9\\
43 9\\
44 9\\
45 9\\
46 9\\
47 9\\
48 9\\
49 9\\
50 9\\
51 9\\
52 9\\
53 9\\
54 9\\
55 9\\
56 9\\
57 9\\
58 9\\
59 9\\
60 9\\
};
\addplot [color=cyellow, line width=2.2pt, forget plot]
  table[row sep=crcr]{%
11 10\\
12 10\\
13 10\\
14 10\\
15 10\\
16 10\\
17 10\\
18 10\\
19 10\\
20 10\\
21 10\\
22 10\\
23 10\\
24 10\\
25 10\\
26 10\\
27 10\\
28 10\\
29 10\\
30 10\\
31 10\\
32 10\\
33 10\\
34 10\\
35 10\\
36 10\\
37 10\\
38 10\\
39 10\\
40 10\\
41 10\\
42 10\\
43 10\\
44 10\\
45 10\\
46 10\\
47 10\\
48 10\\
49 10\\
50 10\\
51 10\\
52 10\\
53 10\\
54 10\\
55 10\\
56 10\\
57 10\\
58 10\\
59 10\\
60 10\\
};
\addplot [color=cpurple, line width=2.2pt, forget plot]
  table[row sep=crcr]{%
19 11\\
20 11\\
21 11\\
22 11\\
23 11\\
24 11\\
25 11\\
26 11\\
27 11\\
28 11\\
29 11\\
30 11\\
31 11\\
32 11\\
33 11\\
34 11\\
35 11\\
36 11\\
37 11\\
38 11\\
39 11\\
40 11\\
41 11\\
42 11\\
43 11\\
44 11\\
45 11\\
46 11\\
47 11\\
48 11\\
49 11\\
50 11\\
51 11\\
52 11\\
53 11\\
};
\addplot [color=cgreen, line width=2.2pt, forget plot]
  table[row sep=crcr]{%
20 12\\
21 12\\
22 12\\
23 12\\
24 12\\
25 12\\
26 12\\
27 12\\
28 12\\
29 12\\
30 12\\
31 12\\
32 12\\
33 12\\
34 12\\
35 12\\
36 12\\
37 12\\
38 12\\
39 12\\
40 12\\
41 12\\
42 12\\
43 12\\
44 12\\
45 12\\
46 12\\
47 12\\
48 12\\
49 12\\
50 12\\
51 12\\
52 12\\
53 12\\
54 12\\
55 12\\
56 12\\
57 12\\
58 12\\
59 12\\
60 12\\
};
\addplot [color=ccyan, line width=2.2pt, forget plot]
  table[row sep=crcr]{%
36 13\\
37 13\\
38 13\\
39 13\\
40 13\\
41 13\\
42 13\\
43 13\\
44 13\\
45 13\\
46 13\\
47 13\\
48 13\\
49 13\\
50 13\\
51 13\\
52 13\\
53 13\\
54 13\\
55 13\\
56 13\\
57 13\\
58 13\\
59 13\\
60 13\\
};
\addplot [color=cmagenta, line width=2.2pt, forget plot]
  table[row sep=crcr]{%
51 14\\
52 14\\
53 14\\
54 14\\
55 14\\
56 14\\
57 14\\
58 14\\
59 14\\
60 14\\
};
\addplot [color=cblue, line width=2.2pt, forget plot]
  table[row sep=crcr]{%
54 15\\
55 15\\
56 15\\
57 15\\
58 15\\
59 15\\
60 15\\
};
\end{axis}
\end{tikzpicture}%
    \vspace{-1.cm}
  \centerline{\small (a)} \medskip
\end{minipage}
\hfill
\begin{minipage}[b]{0.98\linewidth}
  \centering
    \definecolor{cblue}{rgb}{0.066,0.443,0.745}%
\definecolor{corange}{rgb}{0.866,0.329,0.000}%
\definecolor{cyellow}{rgb}{0.929,0.694,0.125}%
\definecolor{cpurple}{rgb}{0.521,0.086,0.819}%
\definecolor{cgreen}{rgb}{0.231,0.666,0.196}%
\definecolor{ccyan}{rgb}{0.184,0.745,0.937}%
\definecolor{cmagenta}{rgb}{0.819,0.015,0.545}%
\begin{tikzpicture}
\begin{axis}[%
width=2.83in, height=1.0in, scale only axis, unbounded coords=jump,
xmin=1, xmax=60, ymin=0.75, ymax=6.5,
xlabel style={font=\footnotesize}, xlabel={Frame index $k$},
ylabel style={font=\footnotesize}, ylabel={Total power [kW]},
axis background/.style={fill=white}, xmajorgrids, ymajorgrids,
legend style={legend cell align=left, align=left, draw=white!15!black, font=\scriptsize, inner sep=1pt, row sep=-1.5pt, at={(0.02,0.98)}, anchor=north west, legend columns=2}, xtick={1,10,20,...,60}
]
\draw[black!45, densely dotted, line width=0.5pt] (axis cs:7,0.75) -- (axis cs:7,6.5);
\draw[black!45, densely dotted, line width=0.5pt] (axis cs:11,0.75) -- (axis cs:11,6.5);
\draw[black!45, densely dotted, line width=0.5pt] (axis cs:19,0.75) -- (axis cs:19,6.5);
\draw[black!45, densely dotted, line width=0.5pt] (axis cs:20,0.75) -- (axis cs:20,6.5);
\draw[black!45, densely dotted, line width=0.5pt] (axis cs:36,0.75) -- (axis cs:36,6.5);
\draw[black!45, densely dotted, line width=0.5pt] (axis cs:51,0.75) -- (axis cs:51,6.5);
\draw[black!45, densely dotted, line width=0.5pt] (axis cs:54,0.75) -- (axis cs:54,6.5);
\addplot [color=corange, line width=1.5pt, mark=square, mark size=2.6pt, mark options={solid, corange}, line width=1.0pt, mark size=2.0pt, mark repeat=6, mark phase=1]
  table[row sep=crcr]{%
1 3.14877\\
2 2.14285\\
3 2.12129\\
4 2.24017\\
5 2.11992\\
6 2.20522\\
7 2.18474\\
8 2.12008\\
9 2.09657\\
10 1.95682\\
11 1.87852\\
12 1.60017\\
13 1.56367\\
14 1.58625\\
15 1.56605\\
16 1.92286\\
17 2.03901\\
18 2.08975\\
19 2.31483\\
20 2.22817\\
21 2.10482\\
22 3.32803\\
23 3.68203\\
24 3.3386\\
25 nan\\
26 nan\\
27 nan\\
28 nan\\
29 nan\\
30 nan\\
31 nan\\
32 nan\\
33 nan\\
34 nan\\
35 nan\\
36 3.82686\\
37 nan\\
38 nan\\
39 nan\\
40 nan\\
41 nan\\
42 nan\\
43 nan\\
44 nan\\
45 nan\\
46 3.04699\\
47 2.20584\\
48 1.96233\\
49 1.81276\\
50 1.61631\\
51 1.76046\\
52 1.42182\\
53 1.26933\\
54 1.37749\\
55 1.21136\\
56 1.10162\\
57 1.15156\\
58 1.07454\\
59 1.1232\\
60 1.42633\\
};
\addlegendentry{Dist., Coop.}
\addplot [color=cmagenta, line width=1.5pt, mark=pentagon, mark size=2.6pt, mark options={solid, cmagenta}, line width=1.0pt, mark size=2.0pt, mark repeat=6, mark phase=1]
  table[row sep=crcr]{%
1 3.14877\\
2 2.14285\\
3 2.12129\\
4 2.10656\\
5 2.10419\\
6 2.11988\\
7 2.3678\\
8 nan\\
9 3.02472\\
10 2.05129\\
11 2.10847\\
12 1.8547\\
13 1.84807\\
14 1.86262\\
15 1.89494\\
16 1.9447\\
17 2.00753\\
18 2.08326\\
19 2.37041\\
20 2.39586\\
21 2.30853\\
22 2.39593\\
23 2.4856\\
24 2.57539\\
25 3.00599\\
26 2.75178\\
27 2.83782\\
28 2.92348\\
29 3.01221\\
30 3.11038\\
31 3.21666\\
32 3.34398\\
33 3.50317\\
34 3.70579\\
35 3.96616\\
36 3.49389\\
37 3.31633\\
38 3.32261\\
39 3.32789\\
40 3.33648\\
41 3.73616\\
42 nan\\
43 nan\\
44 nan\\
45 nan\\
46 3.04699\\
47 2.06978\\
48 1.97002\\
49 1.89123\\
50 1.8251\\
51 1.99145\\
52 1.68636\\
53 1.6571\\
54 1.78498\\
55 1.65043\\
56 1.707\\
57 1.70566\\
58 1.7607\\
59 1.90578\\
60 1.91949\\
};
\addlegendentry{Keep, Coop.}
\addplot [color=ccyan, line width=1.5pt, mark=star, mark size=2.6pt, mark options={solid, ccyan}, densely dashed, line width=1.0pt, mark size=2.0pt, mark repeat=6, mark phase=1]
  table[row sep=crcr]{%
1 2.12976\\
2 1.90729\\
3 1.86541\\
4 1.81924\\
5 1.83674\\
6 1.63431\\
7 1.58964\\
8 1.63187\\
9 1.67565\\
10 1.63593\\
11 1.6214\\
12 1.64925\\
13 1.70442\\
14 1.78627\\
15 1.79443\\
16 1.84214\\
17 1.9646\\
18 2.09329\\
19 2.06046\\
20 2.20806\\
21 2.2717\\
22 2.36382\\
23 2.43928\\
24 2.43022\\
25 2.58939\\
26 2.75932\\
27 2.74629\\
28 2.75233\\
29 2.99359\\
30 3.02084\\
31 2.92527\\
32 2.8409\\
33 2.86589\\
34 2.70854\\
35 2.64012\\
36 2.61989\\
37 2.49175\\
38 2.30017\\
39 2.10332\\
40 2.23415\\
41 2.1943\\
42 2.10623\\
43 2.23503\\
44 2.0993\\
45 1.98809\\
46 1.92702\\
47 1.9139\\
48 1.88145\\
49 1.83654\\
50 1.82558\\
51 1.60228\\
52 1.43878\\
53 1.3896\\
54 1.3205\\
55 1.28547\\
56 1.30564\\
57 1.27802\\
58 1.31514\\
59 1.33168\\
60 1.37852\\
};
\addlegendentry{Joint, memoryless}
\addplot [color=cblue, line width=1.5pt, mark=o, mark size=2.6pt, mark options={solid, cblue}, densely dotted, line width=1.0pt, mark size=2.0pt, mark repeat=6, mark phase=1]
  table[row sep=crcr]{%
1 2.44764\\
2 1.46037\\
3 1.39801\\
4 1.36113\\
5 1.33729\\
6 1.32794\\
7 1.43388\\
8 1.46205\\
9 1.34858\\
10 1.33373\\
11 1.44595\\
12 1.62983\\
13 1.46156\\
14 1.47367\\
15 1.57382\\
16 1.51192\\
17 1.58426\\
18 1.6166\\
19 1.9721\\
20 1.91535\\
21 1.82165\\
22 2.07452\\
23 2.00561\\
24 2.09112\\
25 2.1803\\
26 2.37054\\
27 2.30124\\
28 2.34918\\
29 2.36342\\
30 2.36285\\
31 2.52213\\
32 2.52804\\
33 2.23397\\
34 2.1735\\
35 2.21926\\
36 2.16602\\
37 1.92087\\
38 1.86574\\
39 1.82332\\
40 1.76408\\
41 1.72746\\
42 1.6958\\
43 1.68043\\
44 1.66873\\
45 1.57103\\
46 1.60222\\
47 1.3813\\
48 1.3092\\
49 1.27467\\
50 1.2806\\
51 1.22947\\
52 1.41648\\
53 1.4111\\
54 1.28288\\
55 1.0629\\
56 1.05526\\
57 1.1129\\
58 1.05006\\
59 1.07523\\
60 1.21663\\
};
\addlegendentry{Proposed, look-ahead}
\addplot [color=cblue, line width=1.5pt, mark=diamond, mark size=2.6pt, mark options={solid, cblue}, line width=1.0pt, mark size=2.0pt, mark repeat=6, mark phase=1]
  table[row sep=crcr]{%
1 2.12976\\
2 1.58176\\
3 1.56327\\
4 1.43487\\
5 1.38743\\
6 1.37068\\
7 1.3698\\
8 1.38572\\
9 1.41729\\
10 1.39443\\
11 1.4318\\
12 1.43476\\
13 1.48706\\
14 1.56228\\
15 1.66161\\
16 1.68274\\
17 1.67126\\
18 1.80914\\
19 1.81464\\
20 1.95007\\
21 2.09585\\
22 2.15818\\
23 2.22264\\
24 2.39287\\
25 2.2332\\
26 2.33113\\
27 2.42904\\
28 2.49146\\
29 2.42718\\
30 2.40967\\
31 2.42538\\
32 2.45558\\
33 2.48588\\
34 2.25603\\
35 2.1811\\
36 2.27903\\
37 2.06825\\
38 1.91095\\
39 1.77959\\
40 1.8307\\
41 1.77302\\
42 1.73675\\
43 1.69994\\
44 1.62895\\
45 1.5654\\
46 1.5488\\
47 1.37008\\
48 1.30513\\
49 1.27758\\
50 1.46393\\
51 1.49982\\
52 1.2418\\
53 1.19157\\
54 1.17992\\
55 1.14091\\
56 1.16508\\
57 1.11782\\
58 1.11898\\
59 1.14355\\
60 1.19142\\
};
\addlegendentry{Proposed}
\end{axis}
\end{tikzpicture}%
    \vspace{-1.cm}
  \centerline{\small (b)} \medskip
\end{minipage}
\vspace{-0.4cm}
\caption{Evolution over a trajectory of $60$ frames, i.e., $5$~min, with the parameters of Table~\ref{tab:simu_para}. (a) Membership of the serving set $\mathcal{L}^{(k)}$, with one row per satellite, sorted by the frame in which it first joins. (b) Network-wide power consumption per frame, where vertical dotted lines mark the frames in which the serving set changes.}
\label{time-evolution}
\end{figure}
\subsection{Benchmark Schemes\label{subsec:benchmarks}}
\leavevmode We compare the proposed design with four benchmarks with pre-fixed schedules and with two joint designs that change how Algorithm~\ref{proposed_algo} accounts for the handovers. To isolate the gain of optimizing the schedule jointly with the beamformers, these four benchmarks choose the schedule of each frame by a rule and then optimize the beamformers of all satellites for it by the refinement of Section~\ref{sec_readout}, i.e., under the same model and constraints as the proposed design. The three cooperative rules follow the same two stages: Stage~(i) gives every \ac{ut} one link, without which its \ac{qos} constraint cannot be met, and Stage~(ii) adds cooperative links until each satellite serves $U_{\max}$ \acp{ut}.
\begin{itemize}
    \item \textbf{Distance-based cooperative scheduling:} In Stage~(i), each \ac{ut} is assigned to its nearest satellite with remaining capacity. In Stage~(ii), each satellite adds its nearest \acp{ut} among those it does not serve yet.
    \item \textbf{Correlation-aware cooperative scheduling:} Stage~(i) is distance-based. In Stage~(ii), each satellite adds the \ac{ut} whose largest normalized channel correlation with the \acp{ut} it already serves is the smallest, similar to the scheduling in~\cite{zack2025decentral}. This rule aims at low interference, but it favors \acp{ut} far from those already served and hence tends to select longer links than the distance-based rule.
    \item \textbf{Link-keeping cooperative scheduling:} This rule keeps every previous link whose satellite remains in the serving set, so that these links need no handover, and fills the remaining capacity by the distance-based stages.
    \item \textbf{Distance-based non-cooperative scheduling:} To show the gain of cooperation, each \ac{ut} is served only by the satellite chosen in Stage~(i) of the distance-based rule.
    \item \textbf{Memoryless joint design:} Algorithm~\ref{proposed_algo} runs in every frame as if it were the first one, i.e., with $\delta_{\ell,u}^{(k-1)}=0$ in the design, and thus ignores the links established in the previous frame. Its power and feasibility are evaluated with the actual previous-frame links.
    \item \looseness=-1 \textbf{Look-ahead variant of the proposed design:} Problem~\eqref{ori_prob} counts the whole handover energy of a new link in the frame in which the link is established. Hence, a new link lowers the objective of \eqref{ori_prob} only if it reduces the radiated power of this frame by more than its handover-related power $P_{\mathrm{HO},\ell,u}^{(k)}$, although it can also reduce the radiated power in the later frames in which it serves. The variant therefore counts only a share of this energy in the current frame: it divides $P_{\mathrm{HO},\ell,u}^{(k)}$ in \eqref{eq_P_hat} by $\min\{D_{\ell}^{(k)},N_{\mathrm{la}}\}$, where $D_{\ell}^{(k)}$ is the number of consecutive frames, starting with the $k$-th one, in which the $\ell$-th satellite belongs to the serving set, i.e., the largest number of frames in which the new link can serve. Since the orbits are deterministic, $D_{\ell}^{(k)}$ can be predicted from the ephemeris. The cap $N_{\mathrm{la}}=12$ frames, i.e., $1$~min, accounts for links that are released before their satellite leaves the serving set. Only the coefficients of the handover-related term in \eqref{eq_P_hat} change, so Algorithm~\ref{proposed_algo} applies unchanged, and the power of the variant is computed with \eqref{per_power_model}, i.e., with the whole handover energy, like that of every other scheme.
\end{itemize}

\subsection[Power Consumption over a Trajectory]{Power Consumption over a Trajectory\label{subsec:time}}
\leavevmode We first examine how the power consumption evolves with the handovers. In Fig.~\ref{time-evolution}, we depict the serving set and the network-wide power consumption over a trajectory of $60$ frames, four of which are shown in Fig.~\ref{topology_snapshots}. The serving set changes in 7 frames, each time by one satellite. We make the following observations from Fig.~\ref{time-evolution}(b). First, the proposed design is feasible in all frames. It needs 2.1~kW in the first frame, in which all its 21 links are new, and 1.6~kW in the second, in which it keeps them. Afterwards, its power follows the distance between the \acp{ut} and their nearest serving satellites. Second, on the frames in which both are feasible, the proposed design needs 24\% less power than the distance-based cooperative benchmark, which is infeasible in 20 frames, and 15\% less than the memoryless joint design, which ignores the established links and thus establishes more new links. Third, keeping links does not minimize the power. The link-keeping benchmark establishes no new link from frame 21 to 35, in which the serving set does not change. Over these frames, its links become weaker as the satellites move, and its power rises from 2.3 to 4.0~kW, against at most 2.5~kW for the proposed design. Its schedule is infeasible in frames 8 and 42--45, after which all its links must be re-established. Hence, the handover-related power must be weighed against the radiated power.

The look-ahead variant establishes 30 links in the first frame instead of 21, and thus needs more power there. It then keeps more links, which lower the radiated power, and needs less power than the proposed design in 40 of the remaining 59 frames, and 2\% less over the trajectory.

\begin{figure}[t]
\centering
\definecolor{cblue}{rgb}{0.066,0.443,0.745}%
\definecolor{corange}{rgb}{0.866,0.329,0.000}%
\definecolor{cyellow}{rgb}{0.929,0.694,0.125}%
\definecolor{cpurple}{rgb}{0.521,0.086,0.819}%
\definecolor{cgreen}{rgb}{0.231,0.666,0.196}%
\definecolor{ccyan}{rgb}{0.184,0.745,0.937}%
\definecolor{cmagenta}{rgb}{0.819,0.015,0.545}%
\begin{tikzpicture}
\begin{axis}[hide axis, scale only axis, xmin=0, xmax=1, ymin=0, ymax=1, width=0.5in, height=0.05in,
legend columns=3, legend style={at={(0.5,0.5)}, anchor=center, legend cell align=left, draw=white!15!black, font=\scriptsize, inner sep=1pt, row sep=-1pt, /tikz/every even column/.append style={column sep=2pt}}]
\addlegendimage{color=corange, line width=1.5pt, mark=square, mark size=2.6pt, mark options={solid, corange}, line width=1.2pt, mark size=2.2pt}
\addlegendentry{Dist., Coop.}
\addlegendimage{color=cyellow, line width=1.5pt, mark=triangle*, mark size=2.6pt, mark options={solid, cyellow}, line width=1.2pt, mark size=2.2pt}
\addlegendentry{Corr., Coop.}
\addlegendimage{color=cmagenta, line width=1.5pt, mark=pentagon, mark size=2.6pt, mark options={solid, cmagenta}, line width=1.2pt, mark size=2.2pt}
\addlegendentry{Keep, Coop.}
\addlegendimage{color=cgreen, line width=1.5pt, mark=square, mark size=2.6pt, mark options={solid, cgreen}, line width=1.2pt, mark size=2.2pt}
\addlegendentry{Dist., Non-Coop.}
\addlegendimage{color=ccyan, line width=1.5pt, mark=star, mark size=2.6pt, mark options={solid, ccyan}, densely dashed, line width=1.2pt, mark size=2.2pt}
\addlegendentry{Joint, memoryless}
\addlegendimage{color=cblue, line width=1.5pt, mark=o, mark size=2.6pt, mark options={solid, cblue}, densely dotted, line width=1.2pt, mark size=2.2pt}
\addlegendentry{Proposed, look-ahead}
\addlegendimage{color=cblue, line width=1.5pt, mark=diamond, mark size=2.6pt, mark options={solid, cblue}, line width=1.2pt, mark size=2.2pt}
\addlegendentry{Proposed}
\end{axis}
\end{tikzpicture}%
\vspace{-0.75cm}

\begin{minipage}[b]{0.5\linewidth}
  \centering
    \definecolor{cblue}{rgb}{0.066,0.443,0.745}%
\definecolor{corange}{rgb}{0.866,0.329,0.000}%
\definecolor{cyellow}{rgb}{0.929,0.694,0.125}%
\definecolor{cpurple}{rgb}{0.521,0.086,0.819}%
\definecolor{cgreen}{rgb}{0.231,0.666,0.196}%
\definecolor{ccyan}{rgb}{0.184,0.745,0.937}%
\definecolor{cmagenta}{rgb}{0.819,0.015,0.545}%
\begin{tikzpicture}
\begin{axis}[%
width=1.13in, height=0.88in, scale only axis, unbounded coords=jump,
xmin=0.05, xmax=0.3, ymin=-15.0, ymax=40.0,
xlabel style={font=\footnotesize}, xlabel={Rate requirement\\$\bar{R}_u$ [bps/Hz]},
ylabel style={font=\footnotesize}, ylabel={Power saving [\%]},
axis background/.style={fill=white}, xmajorgrids, ymajorgrids,
legend style={legend cell align=left, align=left, draw=white!15!black, font=\scriptsize, inner sep=1pt, row sep=-1.5pt, draw=none}, xtick={0.05,0.1,0.2,0.3}, minor xtick={0.15,0.25}, xminorgrids, xticklabel style={/pgf/number format/fixed, /pgf/number format/precision=2}, scaled x ticks=false, tick label style={font=\scriptsize}, ylabel shift=-2pt, xlabel style={align=center}, ytick distance=10
]
\addplot [color=corange, line width=1.5pt, mark=square, mark size=2.6pt, mark options={solid, corange}, line width=1.2pt, mark size=2.2pt, forget plot]
  table[row sep=crcr]{%
0.05 9.8417\\
0.1 14.1515\\
0.15 16.6195\\
0.2 16.8525\\
0.25 14.5368\\
0.3 12.717\\
};
\addplot [color=cyellow, line width=1.5pt, mark=triangle*, mark size=2.6pt, mark options={solid, cyellow}, line width=1.2pt, mark size=2.2pt, forget plot]
  table[row sep=crcr]{%
0.05 29.0817\\
0.1 30.5955\\
0.15 32.8112\\
0.2 30.6935\\
0.25 24.7272\\
0.3 nan\\
};
\addplot [color=cmagenta, line width=1.5pt, mark=pentagon, mark size=2.6pt, mark options={solid, cmagenta}, line width=1.2pt, mark size=2.2pt, forget plot]
  table[row sep=crcr]{%
0.05 26.6136\\
0.1 26.5668\\
0.15 23.0161\\
0.2 20.9778\\
0.25 16.7679\\
0.3 19.9428\\
};
\addplot [color=cgreen, line width=1.5pt, mark=square, mark size=2.6pt, mark options={solid, cgreen}, line width=1.2pt, mark size=2.2pt, forget plot]
  table[row sep=crcr]{%
0.05 34.6312\\
0.1 29.4874\\
0.15 nan\\
0.2 nan\\
0.25 nan\\
0.3 nan\\
};
\addplot [color=ccyan, line width=1.5pt, mark=star, mark size=2.6pt, mark options={solid, ccyan}, densely dashed, line width=1.2pt, mark size=2.2pt, forget plot]
  table[row sep=crcr]{%
0.05 16.8101\\
0.1 15.15\\
0.15 14.2365\\
0.2 15.2203\\
0.25 15.9058\\
0.3 17.9348\\
};
\addplot [color=cblue, line width=1.5pt, mark=o, mark size=2.6pt, mark options={solid, cblue}, densely dotted, line width=1.2pt, mark size=2.2pt, forget plot]
  table[row sep=crcr]{%
0.05 -8.38324\\
0.1 -4.45123\\
0.15 -2.32416\\
0.2 -1.08464\\
0.25 -0.665125\\
0.3 -0.0977127\\
};
\addplot [color=cblue, line width=1.5pt, mark=diamond, mark size=2.6pt, mark options={solid, cblue}, line width=1.2pt, mark size=2.2pt, forget plot]
  table[row sep=crcr]{%
0.05 0\\
0.1 0\\
0.15 0\\
0.2 0\\
0.25 0\\
0.3 0\\
};
\end{axis}
\end{tikzpicture}%
    \vspace{-1.cm}
  \centerline{\small (a)} \medskip
\end{minipage}%
\begin{minipage}[b]{0.5\linewidth}
  \centering
    \definecolor{cblue}{rgb}{0.066,0.443,0.745}%
\definecolor{corange}{rgb}{0.866,0.329,0.000}%
\definecolor{cyellow}{rgb}{0.929,0.694,0.125}%
\definecolor{cpurple}{rgb}{0.521,0.086,0.819}%
\definecolor{cgreen}{rgb}{0.231,0.666,0.196}%
\definecolor{ccyan}{rgb}{0.184,0.745,0.937}%
\definecolor{cmagenta}{rgb}{0.819,0.015,0.545}%
\begin{tikzpicture}
\begin{axis}[%
width=1.13in, height=0.88in, scale only axis, unbounded coords=jump,
xmin=0.05, xmax=0.3, ymin=-0.05, ymax=1.05,
xlabel style={font=\footnotesize}, xlabel={Rate requirement\\$\bar{R}_u$ [bps/Hz]},
ylabel style={font=\footnotesize}, ylabel={Feasibility rate},
axis background/.style={fill=white}, xmajorgrids, ymajorgrids,
legend style={legend cell align=left, align=left, draw=white!15!black, font=\scriptsize, inner sep=1pt, row sep=-1.5pt, draw=none}, ytick={0,0.2,0.4,0.6,0.8,1}, xtick={0.05,0.1,0.2,0.3}, minor xtick={0.15,0.25}, xminorgrids, xticklabel style={/pgf/number format/fixed, /pgf/number format/precision=2}, scaled x ticks=false, tick label style={font=\scriptsize}, ylabel shift=-2pt, xlabel style={align=center}
]
\addplot [color=corange, line width=1.5pt, mark=square, mark size=2.6pt, mark options={solid, corange}, line width=1.2pt, mark size=2.2pt, forget plot]
  table[row sep=crcr]{%
0.05 0.996667\\
0.1 0.92\\
0.15 0.734167\\
0.2 0.425833\\
0.25 0.170833\\
0.3 0.0516667\\
};
\addplot [color=cyellow, line width=1.5pt, mark=triangle*, mark size=2.6pt, mark options={solid, cyellow}, line width=1.2pt, mark size=2.2pt, forget plot]
  table[row sep=crcr]{%
0.05 0.981667\\
0.1 0.746667\\
0.15 0.4275\\
0.2 0.164167\\
0.25 0.0525\\
0.3 0.0141667\\
};
\addplot [color=cmagenta, line width=1.5pt, mark=pentagon, mark size=2.6pt, mark options={solid, cmagenta}, line width=1.2pt, mark size=2.2pt, forget plot]
  table[row sep=crcr]{%
0.05 0.981667\\
0.1 0.929167\\
0.15 0.788333\\
0.2 0.5025\\
0.25 0.2425\\
0.3 0.0516667\\
};
\addplot [color=cgreen, line width=1.5pt, mark=square, mark size=2.6pt, mark options={solid, cgreen}, line width=1.2pt, mark size=2.2pt, forget plot]
  table[row sep=crcr]{%
0.05 0.835\\
0.1 0.138333\\
0.15 0.02\\
0.2 0.0108333\\
0.25 0.00583333\\
0.3 0\\
};
\addplot [color=ccyan, line width=1.5pt, mark=star, mark size=2.6pt, mark options={solid, ccyan}, densely dashed, line width=1.2pt, mark size=2.2pt, forget plot]
  table[row sep=crcr]{%
0.05 1\\
0.1 1\\
0.15 0.9975\\
0.2 0.956667\\
0.25 0.676667\\
0.3 0.289167\\
};
\addplot [color=cblue, line width=1.5pt, mark=o, mark size=2.6pt, mark options={solid, cblue}, densely dotted, line width=1.2pt, mark size=2.2pt, forget plot]
  table[row sep=crcr]{%
0.05 1\\
0.1 1\\
0.15 0.999167\\
0.2 0.9975\\
0.25 0.768333\\
0.3 0.379167\\
};
\addplot [color=cblue, line width=1.5pt, mark=diamond, mark size=2.6pt, mark options={solid, cblue}, line width=1.2pt, mark size=2.2pt, forget plot]
  table[row sep=crcr]{%
0.05 1\\
0.1 1\\
0.15 0.9975\\
0.2 0.998333\\
0.25 0.778333\\
0.3 0.378333\\
};
\end{axis}
\end{tikzpicture}%
    \vspace{-1.cm}
  \centerline{\small (b)} \medskip
\end{minipage}
\vspace{-0.4cm}
\caption{Power saving of the proposed design over each scheme on the frames in which both are feasible, (a), and feasibility rate, (b), versus $\bar{R}_u$, where $L=8$, $U_{\max}=4$, $P_{\mathrm{HO}}=40$~dBm, and $\tau_{\mathrm{HO}}=0.1$.}
\label{Rate-QoS}
\end{figure}
\subsection[Impact of the QoS Requirement]{Impact of the QoS Requirement\label{subsec:qos}}
\looseness=-1
We now average the results over $50$ trajectories of $24$ frames. In Fig.~\ref{Rate-QoS}, we report the power saving of the proposed design and the feasibility rate versus the rate requirement $\bar{R}_u$, which ranges from $0.05$ to $0.3$~bps/Hz, i.e., from $12.5$ to $75$~Mbps per \ac{ut}. As shown in Fig.~\ref{Rate-QoS}(b), the proposed design is feasible in at least 99.7\% of the frames up to $0.2$~bps/Hz, whereas the feasibility rate drops to 43\% for the distance-based cooperative benchmark and to 1\% for the non-cooperative benchmark, in which a single satellite serves each \ac{ut} and can rarely deliver the required \ac{sinr} within its power budget. At $0.25$ and $0.3$~bps/Hz, the serving cardinality limits how many satellites can serve each \ac{ut}, and the proposed design is feasible in 78\% and 38\% of the frames, against at most 25\% and 6\% for the pre-fixed schedules. It is also feasible in more frames than the memoryless joint design, which designs every link as new and must deliver the required rate within the data segment alone. As shown in Fig.~\ref{Rate-QoS}(a), the proposed design needs 10--17\% less power than the distance-based cooperative benchmark, 14--18\% less than the memoryless joint design, and at least 16\% less than the other cooperative benchmarks. At $0.1$~bps/Hz, it needs 29\% less than the non-cooperative benchmark, which is feasible in only 14\% of the frames. Hence, joint scheduling and beamforming extends the range of feasible rate requirements, and handover awareness further reduces the power. The look-ahead variant needs 8\% less power than the proposed design at $0.05$~bps/Hz, but about as much at $0.3$~bps/Hz, where both use almost all links.

\begin{figure*}[t]
\centering
\definecolor{cblue}{rgb}{0.066,0.443,0.745}%
\definecolor{corange}{rgb}{0.866,0.329,0.000}%
\definecolor{cyellow}{rgb}{0.929,0.694,0.125}%
\definecolor{cpurple}{rgb}{0.521,0.086,0.819}%
\definecolor{cgreen}{rgb}{0.231,0.666,0.196}%
\definecolor{ccyan}{rgb}{0.184,0.745,0.937}%
\definecolor{cmagenta}{rgb}{0.819,0.015,0.545}%
\begin{tikzpicture}
\begin{axis}[hide axis, scale only axis, xmin=0, xmax=1, ymin=0, ymax=1, width=0.5in, height=0.05in,
legend columns=7, legend style={at={(0.5,0.5)}, anchor=center, legend cell align=left, draw=white!15!black, font=\scriptsize, inner sep=1pt, row sep=-1pt, /tikz/every even column/.append style={column sep=6pt}}]
\addlegendimage{color=corange, line width=1.5pt, mark=square, mark size=2.6pt, mark options={solid, corange}, line width=1.2pt, mark size=2.2pt}
\addlegendentry{Dist., Coop.}
\addlegendimage{color=cyellow, line width=1.5pt, mark=triangle*, mark size=2.6pt, mark options={solid, cyellow}, line width=1.2pt, mark size=2.2pt}
\addlegendentry{Corr., Coop.}
\addlegendimage{color=cmagenta, line width=1.5pt, mark=pentagon, mark size=2.6pt, mark options={solid, cmagenta}, line width=1.2pt, mark size=2.2pt}
\addlegendentry{Keep, Coop.}
\addlegendimage{color=cgreen, line width=1.5pt, mark=square, mark size=2.6pt, mark options={solid, cgreen}, line width=1.2pt, mark size=2.2pt}
\addlegendentry{Dist., Non-Coop.}
\addlegendimage{color=ccyan, line width=1.5pt, mark=star, mark size=2.6pt, mark options={solid, ccyan}, densely dashed, line width=1.2pt, mark size=2.2pt}
\addlegendentry{Joint, memoryless}
\addlegendimage{color=cblue, line width=1.5pt, mark=o, mark size=2.6pt, mark options={solid, cblue}, densely dotted, line width=1.2pt, mark size=2.2pt}
\addlegendentry{Proposed, look-ahead}
\addlegendimage{color=cblue, line width=1.5pt, mark=diamond, mark size=2.6pt, mark options={solid, cblue}, line width=1.2pt, mark size=2.2pt}
\addlegendentry{Proposed}
\end{axis}
\end{tikzpicture}%
\vspace{-0.75cm}

\begin{minipage}[b]{0.245\linewidth}
  \centering
    \definecolor{cblue}{rgb}{0.066,0.443,0.745}%
\definecolor{corange}{rgb}{0.866,0.329,0.000}%
\definecolor{cyellow}{rgb}{0.929,0.694,0.125}%
\definecolor{cpurple}{rgb}{0.521,0.086,0.819}%
\definecolor{cgreen}{rgb}{0.231,0.666,0.196}%
\definecolor{ccyan}{rgb}{0.184,0.745,0.937}%
\definecolor{cmagenta}{rgb}{0.819,0.015,0.545}%
\begin{tikzpicture}
\begin{axis}[%
width=1.18in, height=0.86in, scale only axis, unbounded coords=jump,
xmin=2, xmax=6, ymin=-10.0, ymax=40.0,
xlabel style={font=\footnotesize}, xlabel={Serving cardinality $U_{\max}$},
ylabel style={font=\footnotesize}, ylabel={Power saving [\%]},
axis background/.style={fill=white}, xmajorgrids, ymajorgrids,
legend style={legend cell align=left, align=left, draw=white!15!black, font=\scriptsize, inner sep=1pt, row sep=-1.5pt, draw=none}, xtick={2,3,4,5,6}, tick label style={font=\scriptsize}, ylabel shift=-2pt, ytick distance=10
]
\addplot [color=corange, line width=1.5pt, mark=square, mark size=2.6pt, mark options={solid, corange}, line width=1.2pt, mark size=2.2pt, forget plot]
  table[row sep=crcr]{%
2 nan\\
3 16.5629\\
4 14.1515\\
5 5.50499\\
6 -0.185394\\
};
\addplot [color=cyellow, line width=1.5pt, mark=triangle*, mark size=2.6pt, mark options={solid, cyellow}, line width=1.2pt, mark size=2.2pt, forget plot]
  table[row sep=crcr]{%
2 nan\\
3 25.6629\\
4 30.5955\\
5 30.1266\\
6 27.0486\\
};
\addplot [color=cmagenta, line width=1.5pt, mark=pentagon, mark size=2.6pt, mark options={solid, cmagenta}, line width=1.2pt, mark size=2.2pt, forget plot]
  table[row sep=crcr]{%
2 nan\\
3 19.0457\\
4 26.5668\\
5 21.6401\\
6 7.50378\\
};
\addplot [color=cgreen, line width=1.5pt, mark=square, mark size=2.6pt, mark options={solid, cgreen}, line width=1.2pt, mark size=2.2pt, forget plot]
  table[row sep=crcr]{%
2 nan\\
3 nan\\
4 29.4874\\
5 30.5052\\
6 33.2397\\
};
\addplot [color=ccyan, line width=1.5pt, mark=star, mark size=2.6pt, mark options={solid, ccyan}, densely dashed, line width=1.2pt, mark size=2.2pt, forget plot]
  table[row sep=crcr]{%
2 9.10586\\
3 15.6152\\
4 15.15\\
5 14.1653\\
6 14.6776\\
};
\addplot [color=cblue, line width=1.5pt, mark=o, mark size=2.6pt, mark options={solid, cblue}, densely dotted, line width=1.2pt, mark size=2.2pt, forget plot]
  table[row sep=crcr]{%
2 2.18369\\
3 -2.42438\\
4 -4.45123\\
5 -6.30732\\
6 -7.69255\\
};
\addplot [color=cblue, line width=1.5pt, mark=diamond, mark size=2.6pt, mark options={solid, cblue}, line width=1.2pt, mark size=2.2pt, forget plot]
  table[row sep=crcr]{%
2 0\\
3 0\\
4 0\\
5 0\\
6 0\\
};
\end{axis}
\end{tikzpicture}%
    \vspace{-1.cm}
  \centerline{\small (a)} \medskip
\end{minipage}%
\begin{minipage}[b]{0.245\linewidth}
  \centering
    \definecolor{cblue}{rgb}{0.066,0.443,0.745}%
\definecolor{corange}{rgb}{0.866,0.329,0.000}%
\definecolor{cyellow}{rgb}{0.929,0.694,0.125}%
\definecolor{cpurple}{rgb}{0.521,0.086,0.819}%
\definecolor{cgreen}{rgb}{0.231,0.666,0.196}%
\definecolor{ccyan}{rgb}{0.184,0.745,0.937}%
\definecolor{cmagenta}{rgb}{0.819,0.015,0.545}%
\begin{tikzpicture}
\begin{axis}[%
width=1.18in, height=0.86in, scale only axis, unbounded coords=jump,
xmin=2, xmax=6, ymin=-0.05, ymax=1.05,
xlabel style={font=\footnotesize}, xlabel={Serving cardinality $U_{\max}$},
ylabel style={font=\footnotesize}, ylabel={Feasibility rate},
axis background/.style={fill=white}, xmajorgrids, ymajorgrids,
legend style={legend cell align=left, align=left, draw=white!15!black, font=\scriptsize, inner sep=1pt, row sep=-1.5pt, draw=none}, ytick={0,0.2,0.4,0.6,0.8,1}, xtick={2,3,4,5,6}, tick label style={font=\scriptsize}, ylabel shift=-2pt
]
\addplot [color=corange, line width=1.5pt, mark=square, mark size=2.6pt, mark options={solid, corange}, line width=1.2pt, mark size=2.2pt, forget plot]
  table[row sep=crcr]{%
2 0\\
3 0.44\\
4 0.92\\
5 0.998333\\
6 1\\
};
\addplot [color=cyellow, line width=1.5pt, mark=triangle*, mark size=2.6pt, mark options={solid, cyellow}, line width=1.2pt, mark size=2.2pt, forget plot]
  table[row sep=crcr]{%
2 0.00333333\\
3 0.226667\\
4 0.746667\\
5 0.935\\
6 0.985\\
};
\addplot [color=cmagenta, line width=1.5pt, mark=pentagon, mark size=2.6pt, mark options={solid, cmagenta}, line width=1.2pt, mark size=2.2pt, forget plot]
  table[row sep=crcr]{%
2 0\\
3 0.54\\
4 0.929167\\
5 0.989167\\
6 1\\
};
\addplot [color=cgreen, line width=1.5pt, mark=square, mark size=2.6pt, mark options={solid, cgreen}, line width=1.2pt, mark size=2.2pt, forget plot]
  table[row sep=crcr]{%
2 0\\
3 0.0141667\\
4 0.138333\\
5 0.2825\\
6 0.3325\\
};
\addplot [color=ccyan, line width=1.5pt, mark=star, mark size=2.6pt, mark options={solid, ccyan}, densely dashed, line width=1.2pt, mark size=2.2pt, forget plot]
  table[row sep=crcr]{%
2 0.0525\\
3 0.886667\\
4 1\\
5 1\\
6 1\\
};
\addplot [color=cblue, line width=1.5pt, mark=o, mark size=2.6pt, mark options={solid, cblue}, densely dotted, line width=1.2pt, mark size=2.2pt, forget plot]
  table[row sep=crcr]{%
2 0.0558333\\
3 0.9475\\
4 1\\
5 1\\
6 1\\
};
\addplot [color=cblue, line width=1.5pt, mark=diamond, mark size=2.6pt, mark options={solid, cblue}, line width=1.2pt, mark size=2.2pt, forget plot]
  table[row sep=crcr]{%
2 0.0758333\\
3 0.938333\\
4 1\\
5 1\\
6 1\\
};
\end{axis}
\end{tikzpicture}%
    \vspace{-1.cm}
  \centerline{\small (b)} \medskip
\end{minipage}%
\begin{minipage}[b]{0.245\linewidth}
  \centering
    \definecolor{cblue}{rgb}{0.066,0.443,0.745}%
\definecolor{corange}{rgb}{0.866,0.329,0.000}%
\definecolor{cyellow}{rgb}{0.929,0.694,0.125}%
\definecolor{cpurple}{rgb}{0.521,0.086,0.819}%
\definecolor{cgreen}{rgb}{0.231,0.666,0.196}%
\definecolor{ccyan}{rgb}{0.184,0.745,0.937}%
\definecolor{cmagenta}{rgb}{0.819,0.015,0.545}%
\begin{tikzpicture}
\begin{axis}[%
width=1.18in, height=0.86in, scale only axis, unbounded coords=jump,
xmin=4, xmax=10, ymin=-10.0, ymax=40.0,
xlabel style={font=\footnotesize}, xlabel={Number of satellites $L$},
ylabel style={font=\footnotesize}, ylabel={Power saving [\%]},
axis background/.style={fill=white}, xmajorgrids, ymajorgrids,
legend style={legend cell align=left, align=left, draw=white!15!black, font=\scriptsize, inner sep=1pt, row sep=-1.5pt, draw=none}, xtick={4,6,8,10}, tick label style={font=\scriptsize}, ylabel shift=-2pt, ytick distance=10
]
\addplot [color=corange, line width=1.5pt, mark=square, mark size=2.6pt, mark options={solid, corange}, line width=1.2pt, mark size=2.2pt, forget plot]
  table[row sep=crcr]{%
4 14.6299\\
6 15.3449\\
8 14.1515\\
10 12.3406\\
};
\addplot [color=cyellow, line width=1.5pt, mark=triangle*, mark size=2.6pt, mark options={solid, cyellow}, line width=1.2pt, mark size=2.2pt, forget plot]
  table[row sep=crcr]{%
4 22.0379\\
6 29.5592\\
8 30.5955\\
10 31.2509\\
};
\addplot [color=cmagenta, line width=1.5pt, mark=pentagon, mark size=2.6pt, mark options={solid, cmagenta}, line width=1.2pt, mark size=2.2pt, forget plot]
  table[row sep=crcr]{%
4 16.1235\\
6 22.6126\\
8 26.5668\\
10 27.2487\\
};
\addplot [color=cgreen, line width=1.5pt, mark=square, mark size=2.6pt, mark options={solid, cgreen}, line width=1.2pt, mark size=2.2pt, forget plot]
  table[row sep=crcr]{%
4 23.8517\\
6 28.6876\\
8 29.4874\\
10 27.8423\\
};
\addplot [color=ccyan, line width=1.5pt, mark=star, mark size=2.6pt, mark options={solid, ccyan}, densely dashed, line width=1.2pt, mark size=2.2pt, forget plot]
  table[row sep=crcr]{%
4 10.8218\\
6 12.9217\\
8 15.15\\
10 16.2765\\
};
\addplot [color=cblue, line width=1.5pt, mark=o, mark size=2.6pt, mark options={solid, cblue}, densely dotted, line width=1.2pt, mark size=2.2pt, forget plot]
  table[row sep=crcr]{%
4 -0.0567278\\
6 -1.64558\\
8 -4.45123\\
10 -7.23076\\
};
\addplot [color=cblue, line width=1.5pt, mark=diamond, mark size=2.6pt, mark options={solid, cblue}, line width=1.2pt, mark size=2.2pt, forget plot]
  table[row sep=crcr]{%
4 0\\
6 0\\
8 0\\
10 0\\
};
\end{axis}
\end{tikzpicture}%
    \vspace{-1.cm}
  \centerline{\small (c)} \medskip
\end{minipage}%
\begin{minipage}[b]{0.245\linewidth}
  \centering
    \definecolor{cblue}{rgb}{0.066,0.443,0.745}%
\definecolor{corange}{rgb}{0.866,0.329,0.000}%
\definecolor{cyellow}{rgb}{0.929,0.694,0.125}%
\definecolor{cpurple}{rgb}{0.521,0.086,0.819}%
\definecolor{cgreen}{rgb}{0.231,0.666,0.196}%
\definecolor{ccyan}{rgb}{0.184,0.745,0.937}%
\definecolor{cmagenta}{rgb}{0.819,0.015,0.545}%
\begin{tikzpicture}
\begin{axis}[%
width=1.18in, height=0.86in, scale only axis, unbounded coords=jump,
xmin=4, xmax=10, ymin=-0.05, ymax=1.05,
xlabel style={font=\footnotesize}, xlabel={Number of satellites $L$},
ylabel style={font=\footnotesize}, ylabel={Feasibility rate},
axis background/.style={fill=white}, xmajorgrids, ymajorgrids,
legend style={legend cell align=left, align=left, draw=white!15!black, font=\scriptsize, inner sep=1pt, row sep=-1.5pt, draw=none}, ytick={0,0.2,0.4,0.6,0.8,1}, xtick={4,6,8,10}, tick label style={font=\scriptsize}, ylabel shift=-2pt
]
\addplot [color=corange, line width=1.5pt, mark=square, mark size=2.6pt, mark options={solid, corange}, line width=1.2pt, mark size=2.2pt, forget plot]
  table[row sep=crcr]{%
4 0.301667\\
6 0.804167\\
8 0.92\\
10 0.935833\\
};
\addplot [color=cyellow, line width=1.5pt, mark=triangle*, mark size=2.6pt, mark options={solid, cyellow}, line width=1.2pt, mark size=2.2pt, forget plot]
  table[row sep=crcr]{%
4 0.2725\\
6 0.643333\\
8 0.746667\\
10 0.8025\\
};
\addplot [color=cmagenta, line width=1.5pt, mark=pentagon, mark size=2.6pt, mark options={solid, cmagenta}, line width=1.2pt, mark size=2.2pt, forget plot]
  table[row sep=crcr]{%
4 0.391667\\
6 0.828333\\
8 0.929167\\
10 0.955\\
};
\addplot [color=cgreen, line width=1.5pt, mark=square, mark size=2.6pt, mark options={solid, cgreen}, line width=1.2pt, mark size=2.2pt, forget plot]
  table[row sep=crcr]{%
4 0.105\\
6 0.13\\
8 0.138333\\
10 0.135833\\
};
\addplot [color=ccyan, line width=1.5pt, mark=star, mark size=2.6pt, mark options={solid, ccyan}, densely dashed, line width=1.2pt, mark size=2.2pt, forget plot]
  table[row sep=crcr]{%
4 0.709167\\
6 0.988333\\
8 1\\
10 1\\
};
\addplot [color=cblue, line width=1.5pt, mark=o, mark size=2.6pt, mark options={solid, cblue}, densely dotted, line width=1.2pt, mark size=2.2pt, forget plot]
  table[row sep=crcr]{%
4 0.7725\\
6 0.998333\\
8 1\\
10 1\\
};
\addplot [color=cblue, line width=1.5pt, mark=diamond, mark size=2.6pt, mark options={solid, cblue}, line width=1.2pt, mark size=2.2pt, forget plot]
  table[row sep=crcr]{%
4 0.778333\\
6 0.993333\\
8 1\\
10 1\\
};
\end{axis}
\end{tikzpicture}%
    \vspace{-1.cm}
  \centerline{\small (d)} \medskip
\end{minipage}
\vspace{-0.4cm}
\caption{Power saving of the proposed design over each scheme and feasibility rate versus $U_{\max}$ with $L=8$, (a) and (b), and versus the number of satellites $L$ with $U_{\max}=4$, (c) and (d), where $\bar{R}_u=0.1$~bps/Hz, $P_{\mathrm{HO}}=40$~dBm, and $\tau_{\mathrm{HO}}=0.1$.}
\label{Rate-Links}
\vspace{0.1cm}

\begin{minipage}[b]{0.245\linewidth}
  \centering
    \definecolor{cblue}{rgb}{0.066,0.443,0.745}%
\definecolor{corange}{rgb}{0.866,0.329,0.000}%
\definecolor{cyellow}{rgb}{0.929,0.694,0.125}%
\definecolor{cpurple}{rgb}{0.521,0.086,0.819}%
\definecolor{cgreen}{rgb}{0.231,0.666,0.196}%
\definecolor{ccyan}{rgb}{0.184,0.745,0.937}%
\definecolor{cmagenta}{rgb}{0.819,0.015,0.545}%
\begin{tikzpicture}
\begin{axis}[%
width=1.18in, height=0.86in, scale only axis, unbounded coords=jump,
xmin=30.0, xmax=50.0, ymin=-20.0, ymax=50.0,
xlabel style={font=\footnotesize}, xlabel={Handover-related power\\$P_{\mathrm{HO}}$ [dBm]},
ylabel style={font=\footnotesize}, ylabel={Power saving [\%]},
axis background/.style={fill=white}, xmajorgrids, ymajorgrids,
legend style={legend cell align=left, align=left, draw=white!15!black, font=\scriptsize, inner sep=1pt, row sep=-1.5pt, draw=none}, xtick={30,35,40,45,50}, tick label style={font=\scriptsize}, ylabel shift=-2pt, xlabel style={align=center}, ytick distance=20
]
\addplot [color=corange, line width=1.5pt, mark=square, mark size=2.6pt, mark options={solid, corange}, line width=1.2pt, mark size=2.2pt, forget plot]
  table[row sep=crcr]{%
30 18.2754\\
35 17.3303\\
40 14.1515\\
45 10.7776\\
50 14.2304\\
};
\addplot [color=cyellow, line width=1.5pt, mark=triangle*, mark size=2.6pt, mark options={solid, cyellow}, line width=1.2pt, mark size=2.2pt, forget plot]
  table[row sep=crcr]{%
30 30.5603\\
35 30.5947\\
40 30.5955\\
45 34.4532\\
50 47.1522\\
};
\addplot [color=cmagenta, line width=1.5pt, mark=pentagon, mark size=2.6pt, mark options={solid, cmagenta}, line width=1.2pt, mark size=2.2pt, forget plot]
  table[row sep=crcr]{%
30 30.6674\\
35 29.7517\\
40 26.5668\\
45 22.2496\\
50 22.3126\\
};
\addplot [color=cgreen, line width=1.5pt, mark=square, mark size=2.6pt, mark options={solid, cgreen}, line width=1.2pt, mark size=2.2pt, forget plot]
  table[row sep=crcr]{%
30 34.3723\\
35 32.8265\\
40 29.4874\\
45 26.6072\\
50 20.8252\\
};
\addplot [color=ccyan, line width=1.5pt, mark=star, mark size=2.6pt, mark options={solid, ccyan}, densely dashed, line width=1.2pt, mark size=2.2pt, forget plot]
  table[row sep=crcr]{%
30 11.0109\\
35 11.8105\\
40 15.15\\
45 20.1632\\
50 26.241\\
};
\addplot [color=cblue, line width=1.5pt, mark=o, mark size=2.6pt, mark options={solid, cblue}, densely dotted, line width=1.2pt, mark size=2.2pt, forget plot]
  table[row sep=crcr]{%
30 -0.347292\\
35 -0.818809\\
40 -4.45123\\
45 -9.06874\\
50 -13.4679\\
};
\addplot [color=cblue, line width=1.5pt, mark=diamond, mark size=2.6pt, mark options={solid, cblue}, line width=1.2pt, mark size=2.2pt, forget plot]
  table[row sep=crcr]{%
30 0\\
35 0\\
40 0\\
45 0\\
50 0\\
};
\end{axis}
\end{tikzpicture}%
    \vspace{-1.cm}
  \centerline{\small (a)} \medskip
\end{minipage}%
\begin{minipage}[b]{0.245\linewidth}
  \centering
    \definecolor{cblue}{rgb}{0.066,0.443,0.745}%
\definecolor{corange}{rgb}{0.866,0.329,0.000}%
\definecolor{cyellow}{rgb}{0.929,0.694,0.125}%
\definecolor{cpurple}{rgb}{0.521,0.086,0.819}%
\definecolor{cgreen}{rgb}{0.231,0.666,0.196}%
\definecolor{ccyan}{rgb}{0.184,0.745,0.937}%
\definecolor{cmagenta}{rgb}{0.819,0.015,0.545}%
\begin{tikzpicture}
\begin{axis}[%
width=1.18in, height=0.86in, scale only axis, unbounded coords=jump,
xmin=30.0, xmax=50.0, ymin=0, ymax=6.0,
xlabel style={font=\footnotesize}, xlabel={Handover-related power\\$P_{\mathrm{HO}}$ [dBm]},
ylabel style={font=\footnotesize}, ylabel={New links per frame},
axis background/.style={fill=white}, xmajorgrids, ymajorgrids,
legend style={legend cell align=left, align=left, draw=white!15!black, font=\scriptsize, inner sep=1pt, row sep=-1.5pt, draw=none}, xtick={30,35,40,45,50}, tick label style={font=\scriptsize}, ylabel shift=-2pt, xlabel style={align=center}
]
\addplot [color=corange, line width=1.5pt, mark=square, mark size=2.6pt, mark options={solid, corange}, line width=1.2pt, mark size=2.2pt, forget plot]
  table[row sep=crcr]{%
30 2.10745\\
35 2.10745\\
40 2.10745\\
45 2.10745\\
50 2.10745\\
};
\addplot [color=cyellow, line width=1.5pt, mark=triangle*, mark size=2.6pt, mark options={solid, cyellow}, line width=1.2pt, mark size=2.2pt, forget plot]
  table[row sep=crcr]{%
30 5.29825\\
35 5.29825\\
40 5.29825\\
45 5.29825\\
50 5.29825\\
};
\addplot [color=cmagenta, line width=1.5pt, mark=pentagon, mark size=2.6pt, mark options={solid, cmagenta}, line width=1.2pt, mark size=2.2pt, forget plot]
  table[row sep=crcr]{%
30 1.85354\\
35 1.85354\\
40 1.85354\\
45 1.85354\\
50 1.85354\\
};
\addplot [color=cgreen, line width=1.5pt, mark=square, mark size=2.6pt, mark options={solid, cgreen}, line width=1.2pt, mark size=2.2pt, forget plot]
  table[row sep=crcr]{%
30 2.98765\\
35 2.98765\\
40 2.98765\\
45 2.98765\\
50 2.98765\\
};
\addplot [color=ccyan, line width=1.5pt, mark=star, mark size=2.6pt, mark options={solid, ccyan}, densely dashed, line width=1.2pt, mark size=2.2pt, forget plot]
  table[row sep=crcr]{%
30 3.18\\
35 2.99478\\
40 2.55043\\
45 2.1487\\
50 2.65796\\
};
\addplot [color=cblue, line width=1.5pt, mark=o, mark size=2.6pt, mark options={solid, cblue}, densely dotted, line width=1.2pt, mark size=2.2pt, forget plot]
  table[row sep=crcr]{%
30 2.30783\\
35 2.0713\\
40 1.79565\\
45 1.38174\\
50 1.06783\\
};
\addplot [color=cblue, line width=1.5pt, mark=diamond, mark size=2.6pt, mark options={solid, cblue}, line width=1.2pt, mark size=2.2pt, forget plot]
  table[row sep=crcr]{%
30 1.79913\\
35 1.35652\\
40 1.09217\\
45 0.857391\\
50 0.803136\\
};
\end{axis}
\end{tikzpicture}%
    \vspace{-1.cm}
  \centerline{\small (b)} \medskip
\end{minipage}%
\begin{minipage}[b]{0.245\linewidth}
  \centering
    \definecolor{cblue}{rgb}{0.066,0.443,0.745}%
\definecolor{corange}{rgb}{0.866,0.329,0.000}%
\definecolor{cyellow}{rgb}{0.929,0.694,0.125}%
\definecolor{cpurple}{rgb}{0.521,0.086,0.819}%
\definecolor{cgreen}{rgb}{0.231,0.666,0.196}%
\definecolor{ccyan}{rgb}{0.184,0.745,0.937}%
\definecolor{cmagenta}{rgb}{0.819,0.015,0.545}%
\begin{tikzpicture}
\begin{axis}[%
width=1.18in, height=0.86in, scale only axis, unbounded coords=jump,
xmin=0.05, xmax=0.4, ymin=-20.0, ymax=50.0,
xlabel style={font=\footnotesize}, xlabel={Handover fraction\\$\tau_{\mathrm{HO}}$},
ylabel style={font=\footnotesize}, ylabel={Power saving [\%]},
axis background/.style={fill=white}, xmajorgrids, ymajorgrids,
legend style={legend cell align=left, align=left, draw=white!15!black, font=\scriptsize, inner sep=1pt, row sep=-1.5pt, draw=none}, xtick={0.05,0.1,0.2,0.3,0.4}, xticklabels={$0.05$,,$0.2$,$0.3$,$0.4$}, tick label style={font=\scriptsize}, ylabel shift=-2pt, xlabel style={align=center}, ytick distance=20
]
\addplot [color=corange, line width=1.5pt, mark=square, mark size=2.6pt, mark options={solid, corange}, line width=1.2pt, mark size=2.2pt, forget plot]
  table[row sep=crcr]{%
0.05 14.3176\\
0.1 14.1515\\
0.2 13.517\\
0.3 12.9613\\
0.4 11.8686\\
};
\addplot [color=cyellow, line width=1.5pt, mark=triangle*, mark size=2.6pt, mark options={solid, cyellow}, line width=1.2pt, mark size=2.2pt, forget plot]
  table[row sep=crcr]{%
0.05 30.8282\\
0.1 30.5955\\
0.2 30.3409\\
0.3 29.8925\\
0.4 28.1793\\
};
\addplot [color=cmagenta, line width=1.5pt, mark=pentagon, mark size=2.6pt, mark options={solid, cmagenta}, line width=1.2pt, mark size=2.2pt, forget plot]
  table[row sep=crcr]{%
0.05 26.2862\\
0.1 26.5668\\
0.2 25.7796\\
0.3 24.9527\\
0.4 22.9835\\
};
\addplot [color=cgreen, line width=1.5pt, mark=square, mark size=2.6pt, mark options={solid, cgreen}, line width=1.2pt, mark size=2.2pt, forget plot]
  table[row sep=crcr]{%
0.05 30.0652\\
0.1 29.4874\\
0.2 22.5256\\
0.3 15.6231\\
0.4 nan\\
};
\addplot [color=ccyan, line width=1.5pt, mark=star, mark size=2.6pt, mark options={solid, ccyan}, densely dashed, line width=1.2pt, mark size=2.2pt, forget plot]
  table[row sep=crcr]{%
0.05 10.734\\
0.1 15.15\\
0.2 24.0357\\
0.3 33.1646\\
0.4 42.4913\\
};
\addplot [color=cblue, line width=1.5pt, mark=o, mark size=2.6pt, mark options={solid, cblue}, densely dotted, line width=1.2pt, mark size=2.2pt, forget plot]
  table[row sep=crcr]{%
0.05 -4.74332\\
0.1 -4.45123\\
0.2 -4.50493\\
0.3 -4.61308\\
0.4 -4.92238\\
};
\addplot [color=cblue, line width=1.5pt, mark=diamond, mark size=2.6pt, mark options={solid, cblue}, line width=1.2pt, mark size=2.2pt, forget plot]
  table[row sep=crcr]{%
0.05 0\\
0.1 0\\
0.2 0\\
0.3 0\\
0.4 0\\
};
\end{axis}
\end{tikzpicture}%
    \vspace{-1.cm}
  \centerline{\small (c)} \medskip
\end{minipage}%
\begin{minipage}[b]{0.245\linewidth}
  \centering
    \definecolor{cblue}{rgb}{0.066,0.443,0.745}%
\definecolor{corange}{rgb}{0.866,0.329,0.000}%
\definecolor{cyellow}{rgb}{0.929,0.694,0.125}%
\definecolor{cpurple}{rgb}{0.521,0.086,0.819}%
\definecolor{cgreen}{rgb}{0.231,0.666,0.196}%
\definecolor{ccyan}{rgb}{0.184,0.745,0.937}%
\definecolor{cmagenta}{rgb}{0.819,0.015,0.545}%
\begin{tikzpicture}
\begin{axis}[%
width=1.18in, height=0.86in, scale only axis, unbounded coords=jump,
xmin=0.05, xmax=0.4, ymin=-0.05, ymax=1.05,
xlabel style={font=\footnotesize}, xlabel={Handover fraction\\$\tau_{\mathrm{HO}}$},
ylabel style={font=\footnotesize}, ylabel={Feasibility rate},
axis background/.style={fill=white}, xmajorgrids, ymajorgrids,
legend style={legend cell align=left, align=left, draw=white!15!black, font=\scriptsize, inner sep=1pt, row sep=-1.5pt, draw=none}, ytick={0,0.2,0.4,0.6,0.8,1}, xtick={0.05,0.1,0.2,0.3,0.4}, xticklabels={$0.05$,,$0.2$,$0.3$,$0.4$}, tick label style={font=\scriptsize}, ylabel shift=-2pt, xlabel style={align=center}
]
\addplot [color=corange, line width=1.5pt, mark=square, mark size=2.6pt, mark options={solid, corange}, line width=1.2pt, mark size=2.2pt, forget plot]
  table[row sep=crcr]{%
0.05 0.924167\\
0.1 0.92\\
0.2 0.9125\\
0.3 0.883333\\
0.4 0.8375\\
};
\addplot [color=cyellow, line width=1.5pt, mark=triangle*, mark size=2.6pt, mark options={solid, cyellow}, line width=1.2pt, mark size=2.2pt, forget plot]
  table[row sep=crcr]{%
0.05 0.765833\\
0.1 0.746667\\
0.2 0.698333\\
0.3 0.619167\\
0.4 0.51\\
};
\addplot [color=cmagenta, line width=1.5pt, mark=pentagon, mark size=2.6pt, mark options={solid, cmagenta}, line width=1.2pt, mark size=2.2pt, forget plot]
  table[row sep=crcr]{%
0.05 0.935833\\
0.1 0.929167\\
0.2 0.925\\
0.3 0.91\\
0.4 0.856667\\
};
\addplot [color=cgreen, line width=1.5pt, mark=square, mark size=2.6pt, mark options={solid, cgreen}, line width=1.2pt, mark size=2.2pt, forget plot]
  table[row sep=crcr]{%
0.05 0.163333\\
0.1 0.138333\\
0.2 0.0816667\\
0.3 0.0483333\\
0.4 0.0241667\\
};
\addplot [color=ccyan, line width=1.5pt, mark=star, mark size=2.6pt, mark options={solid, ccyan}, densely dashed, line width=1.2pt, mark size=2.2pt, forget plot]
  table[row sep=crcr]{%
0.05 1\\
0.1 1\\
0.2 1\\
0.3 0.999167\\
0.4 0.9975\\
};
\addplot [color=cblue, line width=1.5pt, mark=o, mark size=2.6pt, mark options={solid, cblue}, densely dotted, line width=1.2pt, mark size=2.2pt, forget plot]
  table[row sep=crcr]{%
0.05 1\\
0.1 1\\
0.2 1\\
0.3 1\\
0.4 1\\
};
\addplot [color=cblue, line width=1.5pt, mark=diamond, mark size=2.6pt, mark options={solid, cblue}, line width=1.2pt, mark size=2.2pt, forget plot]
  table[row sep=crcr]{%
0.05 1\\
0.1 1\\
0.2 1\\
0.3 1\\
0.4 1\\
};
\end{axis}
\end{tikzpicture}%
    \vspace{-1.cm}
  \centerline{\small (d)} \medskip
\end{minipage}
\vspace{-0.4cm}
\caption{Impact of the handover-related power $P_{\mathrm{HO}}$ with $\tau_{\mathrm{HO}}=0.1$, (a) and (b), and of the handover fraction $\tau_{\mathrm{HO}}$ with $P_{\mathrm{HO}}=40$~dBm, (c) and (d), where $\bar{R}_u=0.1$~bps/Hz, $L=8$, and $U_{\max}=4$: power saving of the proposed design, (a) and (c), new links per frame after the first, (b), and feasibility rate, (d).}
\label{Handover-Evaluate}
\end{figure*}
\subsection[Impact of the Serving Cardinality and the Number of Satellites]{Impact of the Serving Cardinality and the Number of Satellites\label{subsec:rfc}}
Next, in Fig.~\ref{Rate-Links}, we vary the number of links that the serving set supports, $LU_{\max}$, through the serving cardinality $U_{\max}$ and the number of satellites $L$, and make the following observations. First, when links are scarce, the schedule determines whether the \ac{qos} can be met. With $U_{\max}=3$ or $L=4$, i.e., $24$ and $16$ links for $12$ \acp{ut}, the proposed design is feasible in 94\% and 78\% of the frames, respectively, against at most 54\% and 40\% for the pre-fixed schedules, and it needs 15--26\% less power than these schedules on the frames in which both are feasible. This is because the pre-fixed rules choose the links independently of the beamforming, whereas the proposed design chooses them together with the beamformers, according to the power needed to meet the \ac{qos} of every \ac{ut}. Second, with $U_{\max}=2$, i.e., also $16$ links, every scheme is feasible in at most 8\% of the frames, since the $4$ satellites closest to the \acp{ut} can then serve only $8$ of them, and the others depend on farther satellites, whose links are weaker. Third, as $U_{\max}$ grows from $4$ to $6$, the saving over the distance-based cooperative benchmark drops from 14\% to 6\% and vanishes. This benchmark uses all $LU_{\max}$ links, which lowers the radiated power of its beamformers, whereas the proposed design uses fewer links, e.g., 28 instead of 48 at $U_{\max}=6$, since it counts the whole handover energy of a new link in a single frame. The look-ahead variant uses 43 links there and needs 8\% less power than the proposed design. In contrast, the saving over the distance-based benchmark is still 12\% at $L=10$, since the added links are to farther satellites, and the look-ahead variant needs 7\% less power there. Hence, joint scheduling and beamforming matters most when links are scarce, and the look-ahead variant when links are plentiful.

\subsection[Impact of the Handover-Related Power and Duration]{Impact of the Handover-Related Power and Duration\label{subsec:ho}}
Finally, we study the impact of the handover-related power and duration, which determine the cost of a new link. In Fig.~\ref{Handover-Evaluate}(a) and (b), we vary $P_{\mathrm{HO}}$ from $30$ to $50$~dBm, i.e., $E_{\mathrm{HO}}$ from $5$ to $500$~J, and report the power saving of the proposed design and the number of new links per frame. As expected, the proposed design establishes fewer new links as $P_{\mathrm{HO}}$ grows, whereas the pre-fixed schedules do not depend on $P_{\mathrm{HO}}$. It needs 11--18\% less power than the distance-based cooperative benchmark, 22--31\% less than the link-keeping benchmark, and 11--26\% less than the memoryless joint design. The last saving grows with $P_{\mathrm{HO}}$, since the memoryless design establishes more new links. The look-ahead variant also establishes more new links than the proposed design, since it counts only a share of their handover energy in the current frame. At $40$~dBm, these links add 48~W of handover-related power but save 133~W of radiated power, so the variant needs 4\% less power than the proposed design. This saving grows from 0.3\% at $30$~dBm to 13\% at $50$~dBm, since the handover energy that the proposed design counts in a single frame grows with $P_{\mathrm{HO}}$.

In Fig.~\ref{Handover-Evaluate}(c) and (d), we vary the handover fraction $\tau_{\mathrm{HO}}$ from $0.05$ to $0.4$, i.e., the handover duration from $0.25$ to $2$~s. As shown in Fig.~\ref{Handover-Evaluate}(d), the proposed design is feasible in all frames, whereas the pre-fixed schedules become infeasible in more frames: the feasibility rate drops from 92\% to 84\% for the distance-based cooperative benchmark and from 16\% to 2\% for the non-cooperative benchmark. Moreover, the power of the proposed design grows by only 4\%, whereas that of the memoryless joint design grows by 61\%, so that the saving over this design grows from 11\% to 42\% (Fig.~\ref{Handover-Evaluate}(c)). This is because a longer handover mainly affects the \acp{ut} whose links are all new, which receive no data during the handover-decision subframe (Remark~\ref{rem:continuity}). With cooperation, most \acp{ut} keep a persisting link from one frame to the next, whereas the memoryless joint design treats every link as new in its design, and the non-cooperative benchmark gives each \ac{ut} a single link, which is new whenever the \ac{ut} is handed over. Hence, the handover awareness of the proposed design matters more as handovers become more costly or longer.

\section{Conclusion\label{sec_conc}}
This paper investigated power-efficient downlink transmission in networked \ac{leo} satellite systems by accounting for the cost of handovers in the joint design of cooperative beamforming and user scheduling under statistical \ac{csi}. Based on a two-segment frame structure, we modeled the energy and the transmission time that each new satellite-\ac{ut} link costs, so that the design can weigh the handover cost of a new link against the radiated power that the link saves. We minimized the resulting network-wide power per frame by an iterative algorithm with convex subproblems.

Based on simulations over two-minute trajectories with vessels in the Red Sea, we draw three conclusions. First, joint scheduling and beamforming saves power: at $25$~Mbps per \ac{ut}, the proposed design needs 14\% less power than distance-based cooperative scheduling, and it meets the rate requirements in all frames, against 92\% of the frames for this scheme and 14\% for non-cooperative transmission. Second, the handover-related power must be weighed against the radiated power: the proposed design needs 15\% less power than a joint design that ignores the established links and 27\% less than a schedule that keeps every link while its satellite remains in the serving set. Third, cooperation makes the design robust to the handover duration: most \acp{ut} keep a link that carries data during the handover, so the power of the design grows by only 4\% as the handover duration grows from $0.25$ to $2$~s. A look-ahead variant, which divides the handover energy of a new link by the predicted number of frames, up to 12, in which the link can serve, needs 4\% less power than the proposed design. Overall, joint scheduling and beamforming matters most when the rate requirements are high or the links are scarce, and handover awareness matters most when handovers are costly or long.

Future work may optimize the schedules jointly over several predicted frames, model \ac{ut}-dependent handover durations and imperfect ephemeris-aided compensation, and decentralize the design under limited \ac{isl} capacity and latency.

\bibliographystyle{IEEEtran}
\bibliography{IEEEabrv,mybib}

\end{document}